\documentstyle[aaspp]{article}

\begin{document}

\title{Improving Predictions for Helium Emission Lines}

\author{Robert A. Benjamin \altaffilmark{1}}
\affil{University of Wisconsin-Madison, Dept. of Physics, 1150 University Avenue, Madison, WI 53706; benjamin@wisp.physics.wisc.edu}

\author{Evan D. Skillman}
\affil{Department of Astronomy, University of Minnesota, Minneapolis,MN 55455;
skillman@ast1.spa.umn.edu}

\author{Derck P. Smits}
\affil{Dept Math \& Astronomy, University of South Africa, PO Box 392, Pretoria
0003 South Africa; dps@pc-2029227.unisa.ac.za}

\altaffiltext{1}{Previous address: Minnesota Supercomputer Institute, 1200 
Washington Ave, SE, Minneapolis, MN 55415}

\begin{abstract}

We have combined the detailed He I recombination model of Smits with the
collisional transitions of Sawey \& Berrington in order to produce new
accurate helium emissivities that include the effects of collisional
excitation from both the $2 ^{3} S$ and $2 ^{1} S$ levels. We present
a grid of emissivities for a range of temperature and densities along
with analytical fits and error estimates. These grids eliminate the
necessity of making corrections for collisional enhancements as in the
work of Clegg or Kingdon \& Ferland for lines with upper levels below
n=5.  For densities greater than $n_{e} \approx 10^{6}~cm^{-3}$,
inclusion of collisional excitation from the $2 ^{1} S$ level is also
necessary if accuracies of greater than a few percent are required.

Atomic data for a model atom with 29 levels ($n_{max}=5$) is presented
which matches the recombination model of Smits to within 5\% over the
temperature range $T=5000-20000~{\rm K}$.  Using the algorithm of
Almog \& Netzter, collisional effects are calculated
self-consistently. This model atom will be useful in models of
radiative transfer. A notable feature of this technique is an
algorithm which calculates the ``indirect'' recombination rates, the
recombination to individual levels which go through $n>n_{max}$ first.

Fits accurate to within 1\% are given for the emissivities of the
brightest lines over a restricted range for estimates of primordial
helium abundance. We characterize the analysis uncertainties associated
with uncertainties in temperature, density, fitting functions, and 
input atomic data. We estimate that atomic data uncertainties alone may 
limit abundance estimates to an accuracy of $\sim 1.5 \%$;
systematic errors may be greater than this. This analysis uncertainty
must be incorporated when attempting to make high accuracy estimates of
the helium abundance. For example, in recent determinations of the primordial
helium abundance, uncertainties in the input atomic data have been neglected.

Finally, we compare our theoretical calculations to the measured
strengths of a few dozen helium emission lines in three nebulae, Orion
(NGC 1976) and the planetary nebulae, NGC 6572 and IC 4997.
Incorporation of collisional effects yields noticeable improvements
for some lines, but some notable discrepancies remain.

\end{abstract}

\section{Motivation}

The majority of helium was created in the first few minutes
of the Universe's existence and was first discovered in 1868 (Lockyer
1868) as an emission line at 5876 \AA~ in the spectrum of a solar
prominence.  Since then it has become clear that measuring the
abundance of helium in astrophysical objects can be used to constrain
the parameters of primordial nucleosynthesis (see Walker et al
1991).  Helium abundance is a particularly useful diagnostic, since
unlike deuterium and lithium, its abundance only increases with
time. Measurement of accurate helium abundances in low abundance
objects sets a firm upper limit on the primordial helium abundance,
and extrapolation of the helium abundance down to zero metallicity can
be used to estimate the primordial helium abundance (Peimbert \&
Torres-Peimbert 1974).

While the monotonic change of helium abundance with time is useful,
the variation of the primordial value of the helium abundance with
cosmological parameters is sufficiently small that extremely accurate
measurements of the helium abundance are required to provide useful
cosmological constraints. This requires precise measurements of helium
emission lines and accurate models of the recombination spectrum of
helium and an understanding of the uncertainties in both. Previously
it has been assumed that the uncertainties associated with the
recombination model were significantly less than the observational
uncertainties. However, with observations now able to determine helium
abundance to within $\sim 2\%$ (Skillman et al. 1994), it has become
necessary to revisit this assumption.  Improving the recombination
model and assessing the uncertainties is the focus of this paper. In
addition to being useful for constraining helium abundances, an
accurate recombination model is also useful for providing constraints
on nebular parameters such as optical depths, temperatures, and
extinction.

Numerous authors have calculated the recombination spectrum of neutral helium
(Mathis 1957; Burgess \& Seaton 1960a,b; Pottasch 1961; Robbins 1968;
Robbins 1970; Robbins \& Robinson 1971; Bhatia \& Underhill 1985;
Brocklehurst 1972; Almog \& Netzer 1989; Smits 1991b,1996).  The
calculations over this span have been characterized by improvements in
atomic data, an increase in accuracy as more levels can be accurately
included, and the rooting out of occasional bugs or inaccurate
approximations. The most accurate treatment of emissivities from the
helium atom in nebular environment currently available is that of
Smits (1996). This work used the most accurate calculations of
spontaneous radiative transitions to date (Kono \& Hattori 1984),
which supersedes the compilation of Wiese, Smith \& Glennon (1966). It
also recalculated the radiative recombination rates for several levels
using updated He I photoionization cross sections of Fernley et al
(1987). In addition, it corrects an error in Brocklehurst (1972)
affecting the $2^{3}P-n^{3}S$ series and an error affecting high
temperature recombination rates in Smits (1991b).

With accuracies of a few percent required, it has also been found that
collisional excitation from the metastable $2^{3}S$ plays a
significant role for several bright lines, even at the low densities 
characteristic of extragalactic H II regions.  Twice in the past
(Cox \& Daltabuit 1971; Ferland 1986) it has
been suggested that high values of collisional excitation rates might
force a significant re-evaluation of derived helium abundances,
although in both cases it turned out that the collisional rates were
overestimates. In order to constrain these collisional corrections to
the radiative cascade, quantum calculations of increasing accuracy
have been carried out to determine more exact collisional rates (Berrington
et al. 1985; Berrington \& Kingston 1987; Sawey \& Berrington 1993).

\section{A New Model Helium Atom}

Here, we present calculations of helium emission which combine the
recombination and radiative cascade data of Smits (1996, hereafter
S96) with the collisional rates of Sawey \& Berrington (1993,
hereafter SB93). Using a similar algorithm to that used by Almog \&
Netzer (1989, hereafter AN89), we construct model helium atoms with a
small number of levels which match the calculations of S96 to within
less than 0.2\% for $n_{e} \leq 10^{6}~cm^{-3}$. These calculations
are valuable for two reasons. First, since collisional effects are
completely and self-consistently included in these calculations, it
eliminates the need to estimate collisional correction factors for
emission line intensities, as in Clegg (1987) or Kingdon \& Ferland
(1995, hereafter KF95). And second, the small but accurate model atom
presented here can then be used in a model of radiative transfer 
which will allow these models to be calculated rapidly
without a loss in accuracy for prediction of emission line
intensities.

\subsection{Atomic Data}

The data used here are described in SB93, S96 and references therein,
and are summarized in the Appendix A. The values for a $n_{max}=5$
model helium atom is given in Tables 1-4. In these tables we present
power-law fitting functions for the temperature dependent rates. For
example, the recombination rate to an individual level is given by
$\alpha_{nl}(T)=\alpha_{4} t^{b_{\alpha}}$, where $t \equiv
T/10^{4}~K$.  For all rates, the power-law is chosen such that at
$t=1$ the agreement with our calculations is exact. We also tabulate
$\pm e_{\alpha}$, the maximum percentage error associated with the
fits over the temperature range $t=0.5-2.0$. We also give fits to the
{\it total} Case B recombination rates into the singlet and triplet
ladders.  Table 2 contains fits for the three body recombination rates,
given by $\alpha^{tbr}=\alpha_{4}^{tbr} t^{b_{\alpha^{tbr}}}$, and the
fits to the collisional ionization rate, $C^{ion}=a t^{b}e^{c/t}$.
For temperatures outside this range, for applications which require a
higher degree of accuracy, and for data for larger model atoms,
readers should contact the authors for an extensive grid of rates.

Note that by using fits to the atomic data rather than the original atomic
data themselves, the resulting calculations will be slightly less accurate. 
Over the temperature range $t=0.5-2.0$, the line fluxes derived using the 
data in Tables 1-4 are acccurate to within
9.6\% for $n_{e}< 10^{8}~cm^{-3}$, and accurate to within 6.1\% for 
$n_{e}< 10^{6}~cm^{-3}$. In all results reported here, we use the original
atomic data rather than the fitting functions given in the tables.

Although we have incorporated all of the collisional transitions in
our calculations, numerical experimentation described below has shown
that for densities as high as $n_{e}=10^{8}~cm^{-3}$, only collisional
transitions from $2^{1}S$ and $2^{3}S$ and those with $\Delta n=0$
affect the predicted line fluxes. The effects of turning on all other
collisions changes the line fluxes by less than 0.01 \%. However we
include all collision strengths in Table 4.

We can match the $N \sim 3000$ level calculation of S96 with only
$N=29$ levels (for $n_{max}=5$) using an algorithm which calculates
the ``indirect recombination'' rate, $\alpha^{\prime}(n_{e},T)$, from
levels above $n_{max}$ into the individual levels of our model atom.
These rates are in Table 3.  They have a weak density dependence since
the recombination cascade in the levels above $n_{max}$ is affected by
collisional mixing. Below $n_{e,2} \equiv n_{e}/10^{2}~cm^{-3}=1$, the
indirect recombination rates have negligible density dependence and
are fit by $\alpha^{\prime}=\alpha_{4}^{\prime}
t^{b_{\alpha^{\prime}}}$, while above this density we use
$\alpha^{\prime}=\alpha_{4}^{\prime} t^{b_{\alpha^{\prime}}}
n_{e,2}^{c_{\alpha^{\prime}}}$. Holding the density fixed at
$n_{e}=10^{2}~cm^{-3}$, the maximum percentage error associated with
this fit over the temperature range $t=0.5-2.0$ is
$e_{\alpha^{\prime}}(n_{e})$. Holding the temperature fixed at $t=1$,
the maximum error over the density range $n_{e}=10^{2}
-10^{6}~cm^{-3}$ is $e_{\alpha^{\prime}}(T)$.  For the $n_{max}=5$
atom presented here, the indirect recombination rate to most levels
ranges from 20 \% to 50 \% of the sum of the indirect and direct
rates.

The indirect recombination rates account for recombination that passes
through levels greater than $n_{max}$, where $n_{max}$ is the upper
cutoff for the model atom. In previous works, such as AN89 and Cota
(1987), the upper levels have been replaced by one or two
``fictional'' level with a set of $A$-values from these levels.
However, no attempts were made to ``tune'' the A-values from
the ``fictional'' levels such that the population of given level with
$n < n_{max}$ \underline{exactly} matches the value that it would have
had in the general case.  For the purposes here,
we needed a more accurate algorithm.

We calculate the indirect recombination as follows. Following
Osterbrock (1989), we define the probability and cascade matrices,
${\bf P}$ and ${\bf C}$ where

\begin{equation}
P_{i,k} \equiv A_{i,k}/\sum_{E_{i}>E_{j}} A_{i,j}~,
\end{equation} 

is the probability that the population of level $i$ will be followed by 
a radiative decay to level $k$. The elements of the cascade matrix are

\begin{equation}
C_{i,k} \equiv \sum_{E_{i}>E_{j}>E_{k}} C_{i,j} P_{j,k} ~,
\end{equation}

which gives the probability that the population of $i$ will eventually yield 
a transition to level $k$ via all intermediate routes $j$, where
it is assumed that $C_{i,i} \equiv 1$.

For a pure radiative cascade, the population $n_{k}$ of a level $k$ is
set by balancing the rates for population and de-population. The depopulation
rate is the sum all possible radiative decays, i.e. $\sum_{l}
A_{k,l}$ where $E_{k} >E_{l}$. The population rate consists of three
terms: (1) direct recombination rate to level $k$, $n_{e}\alpha_{k}$,
(2) direct recombination to all included upper levels $i$ followed by
a cascade to level ${k}$, $\sum_{i} n_{e}\alpha_{i}C_{i,k}$, and (3)
the indirect recombination, $n_{e}\alpha_{k}^{\prime}$.

If $n_{k}$ is the population of level $k$ as derived by the full
calculation of S96, one can solve for the indirect
recombination rate

\begin{equation}
n_{e}n_{He^{+}} \alpha_{k}^{\prime}=n_{k} \sum_{l}A_{k,l}- n_{e}n_{He^{+}}\sum_{i}(\alpha_{i}+\alpha_{i}^{\prime})C_{i,k} ~.
\end{equation}

This relation can be solved for successively from the highest energy
level to the lowest. This formulation allows us to match the level
populations of S96 exactly with a finite number of levels, so long as
collisional processes are negligible in determining the level
populations.  Note that the sum of all indirect and direct
recombination rates should equal the total recombination rate; we have
checked to verify this and found it to be true within 2\% for $n_{e}
\leq 10^{6}$, and 10 \% for $n_{e}=10^{8}~cm^{-3}$.  We have also used
these indirect recombination rates with the data otherwise identical
to those of S96 and found that the level populations from our model
atom agree with the calculation of S96 to better than 0.2\% for $n_{e}
\leq 10^{6}$. Note that in this test, although our data is identical
to that of Smits (1996), the numerical method used to calculate
the level populations, described in the next section, is independent.

It should be noted that this ``indirect'' recombination rate is
different from the ``effective'' recombination rate for a given level,
as in Osterbrock (1989), since the effective recombination rate to a
given level will also include the cascade from levels $n \leq 5$, as
well as the direct and indirect recombination. The effective
recombination rate may be calculated using the emissivities in Table 5-7
together with the A-values in Table 4.

\section{Emissivities}

Because of the coupling of the singlet and triplet levels via collisional
routes, we use a matrix inversion routine to solve for the level populations
 and the resultant emissivities. The equation of statistical equilibrium for
a level $k$, 

\begin{eqnarray} 
n_{e}n_{He^{+}}\left[\alpha_{k}(T)+\alpha_{k}^{\prime}(T,n_{e})+
                               n_{e}\alpha_{k}^{tbr}(T)   \right]+
\sum_{E_{i}>E_{k}}n_{i}A_{i,k}+ \nonumber \\
\sum_{X}\sum_{j}n_{X}n_{j}q_{j,k}^{X}= 
n_{k} \left[ n_{e}C^{ion}_{k}+
\sum_{E_{k}>E_{l}}A_{k,l}+ 
\sum_{X} \sum_{j} n_{X}q_{k,j}^{X}\right] ~,
\end{eqnarray} \label{eq-stateq}

can be used to calculate the populations for $k_{max}=n_{max}(n_{max}+1)-1$
levels, where $n_{max}$ is the maximum principal quantum number for
the model atom. The collisional transitions with species
$X=e,p,He^{+}$ and radiative decay from $2^{3}S$ to $1^{1}S$ can couple the
singlet ($S=0,g_{S}=1$) and triplet ($S=1,g_{S}=3$) recombination
ladders, so that level populations for both ladders must be solved
simultaneously. The index of an individual level characterized by
quantum numbers $n$, $l$, and $S$ is $K=2 [n (n-1)/2 +l]+(1-S)$.

These equations are solved in the Case B limit, in which photons
associated with all permitted radiative transitions to $1^{1}S$ are
assumed to be reabsorbed. In Case B, all electric dipole transitions
to $1^{1}S$ and $\alpha_{1}$ are set to zero.

Because of the collisional coupling between the singlet and triplet ladders,
we can not apply a simple iterative scheme to determine the level populations 
as in Brocklehurst (1972) or Smits(1991a). Instead, we use the same algorithm
as AN89. This involves expressing the rate equations in matrix form: ${\bf R
\cdot n = \alpha}$, where

\begin{equation}
\left( \begin{array}{ccccc}
-\Gamma+n_{e}{\cal Q}_{1}+n_{e}C^{ion}_{1}                 &  A_{2,1}+n_{e}q_{2,1}                              & 
 \ldots & A_{k_{max},1}+q_{k_{max},1} \\
n_{e}q_{1,2}         & -({\cal A}_{2}+n_{e}{\cal Q}_{2}+n_{e}C^{ion}_{2}) & 
 \ldots & A_{k_{max},2}+q_{k_{max},2} \\
\vdots               &                                                    &
\ddots  & \vdots        \\
n_{e}q_{1,k_{max}}         & n_{e}q_{2,k_{max}} & \ldots & -({\cal A}_{k_{max}}+n_{e}{\cal Q}_{k_{max}}+n_{e}C^{ion}_{k_{max}})
\end{array} \right)
{\bf n}
=
-n_{e}n_{He^{+}} {\bf \alpha^{tot}}
\end{equation}

${\bf n}=(n_{1}, \ldots, n_{k_{max}})$ is the vector of level populations
and ${\bf \alpha^{tot}}=(\alpha_{1}^{tot}, \ldots, \alpha_{k_{max}})$
is the vector of recombination rates, where
$\alpha^{tot}_{k}=\alpha_{k}+\alpha_{k}^{\prime}+n_{e}\alpha_{k}^{tbr}$.

In the rate matrix, the collisional rate coefficient is $q_{k,j}=\sum_{X} (n_{X}/n_{e}) q_{k,j}^{X}$, and
$\Gamma$ ($s^{-1}$) is the photoionization rate from the ground state.
Here it is assumed that the levels have been rearranged such that the index
increases monotonically with energy, i.e., $k_{2}>k_{1}$ implies
$E_{k_{1}} > E_{k_{2}}$. The diagonal elements of ${\bf R}$ are the
depopulation rates from a given level due to radiative decays (${\cal
A}=\sum_{l}A_{k,l}$), collisional transitions (${\cal
Q}=\sum_{X}\sum_{j}n_{X}q_{k,j}$), and collisional ionization
($n_{e}C^{ion}$). The entries above the diagonal are the decays, the
entries below are the excitations, and the right hand matrix contains
the population rates of each level due to recombination. This
matrix will be sparse, since many of the elements are zero. For example,
singlet and triplet energy levels are only coupled by three radiative
decay routes;  $A_{k_{1},k_{2}}$'s are
zero unless $l_{k_{1}}=l_{k_{2}} \pm 1$, and so on.

Inversion of ${\bf R}$ using a standard numerical routine for matrix
inversion (Press et al. 1986) allows one to solve for the matrix
of level populations $n$. If there were no coupling between singlet
and triplets, the matrix could be broken into two independent
submatrices and solved faster. In the limit of a pure radiative
cascade ($n_{e} \rightarrow 0$) with no forbidden transitions, the
matrix will become singular, since there are no depopulation
mechanisms for the $2^{1}S$ or $2^{3}S$ levels. In this limit, the
rows and columns in the above matrix equation corresponding to these
levels must be removed.

Using these level populations, we calculate the line emissivities.
Tables 5-7 give the emission coefficient, $4 \pi
j_{line}/n_{e}n_{He+}$, for the He I emission line at 4471 \AA~ for a
range of temperature and densities. For convenience we also include
the He II 4686 \AA~ line, and hydrogen $H\beta$ and $Br\gamma$ lines
calculated by Storey \& Hummer (1995). We give the ratio of
emissivities for all emission lines with $j_{line}/j_{4471}>10^{-2}$.
However, this table only includes lines originating from levels $n
\leq 5$ with $\lambda < 2.5 \mu m$.  The wavelengths of the lines are
those calculated using the energy levels of Table 1, and will differ
from wavelengths observed in air, e.g., the 4471 \AA~ line in air is
reported as 4473 \AA, the vacuum wavelength. The formula for
conversion from between air and vacuum can be found in Morton (1991).
Where ambiguities arise, the quantum numbers corresponding to a given
wavelength can be obtained from Table 4.

Lines whose fluxes are thought to be uncertain to more than 5\% due to
the lack of collisional data from the $2^{1}S$ and $2^{3}S$ levels to
levels with $n \geq 5$ are marked. See \S3.1 for further discussion.
The last four columns in Tables 5-7 contain a fit to the emission
coefficients of the form $4 \pi j_{line}/n_{e}n_{He+}=a_{j} t^{b_{j}}
e^{c_{j}/t}$, where the maximum percentage difference between the
fitting function and the calculated results is $e_{j}$. A grid with an
extended density and temperature range, finer resolution, and more
transitions, is available from the authors.

\subsection{Collisional effects}

In estimating helium abundances it has become standard to use the
recombination cascade calculations of Brocklehurst (1972) or S96 and
apply a correction for collisional enhancement of these lines. Since
all of the rates out of the $2^{3}S$ level are low, substantial
populations can build up in this level, and collisional transitions from
$2^{3}S$ to
upper levels are significant. KF95 have most
recently estimated these effects using the same atomic data set as
this calculation. In Table 8, we compare our calculated population
fraction in the $2^{3}S$ levels with those
derived by KF95, where they used the formula

\begin{equation}
\frac{n(2^{3}S)}{n_{He^{+}}}=\frac{n_{e}\alpha_{B}^{3}}{A_{21}+n_{e}q_{tot}}~,
\end{equation}

where $\alpha_{B}^{3}$ is the case B recombination into the triplet
levels, $A_{21}$ is the spontaneous decay rate to $1 ^{1}S$ and
$q_{tot}$ is the sum of all collisional transition rates (both
excitation and ionization) from $2 ^{3}S$ into the singlet ladder.
We show cases for $t=2$ since higher temperatures give larger
Boltzmann factors, emphasizing the collisional enhancement.

For the low density regime, $n_{e} \leq 10^{2}~cm^{-3}$, where $A_{21}
n_{e}q_{tot}$, our calculated $2 ^{3}S$ population is systematically
3.3\% higher than that of KF95. This is completely attributable to the
difference in the case B triplet recombination rate,
$\alpha_{B}^{3}$. KF95 used a fit to the recombination rate of
Osterbrock (1989) which underestimated this rate by 3.3\%. Although
S96 did not report the total recombination rate, his value matched
that of Brocklehurst (1972) nearly exactly.

For higher densities, where the collisional term becomes important,
the remaining differences are due to two factors: First, KF95 use a
power-law fit to the collisional rate, $q_{tot}=3.61 \times 10^{-8}
t^{0.5}$, which differs from our value by only 1\% at $t=1$, but
exceeds our rate by 40 \% at $t=0.5$ and 19\% at $t=2$. This
leads KF95 to underestimate the $2^{3}S$ population. Hoever, these 
differences do not translate directly into emissivity differences.
 Second, as the
densities increase, a non-negligible population builds up in the
$2^{1}S$ level, and electrons are transfered back from the singlet
into the triplet ladder. For $t=1$, this effect increases the $2^{3}S$
population from what one calculates using the above equation by 6\%
for $n_{e}=10^{6}~cm^{-3}$ and 94\% for $n_{e}=10^{8}~cm^{-3}$.  In
high density environments, therefore, it is necessary to calculate the
singlet and triplet ladders simultaneously, as emphasized by AN89.

Table 8 also shows our calculation for the ratio of
collisional-to-recombination contributions for selected emission lines
and a comparison to the values calculated from KF95. Apart from
propagating the differences for $n(2^{3}S)$, the remaining differences
are due to the fact that KF95 included collisional excitation to $n=5$
for selected lines.  For emission lines originating from $n_{u} \leq
4$, this contribution is less than 5\%, for
$n_{e} \leq 10^{6}~cm^{-3}$, but for lines from $n_{u}=5$, such as
4026 \AA~ or 4387 \AA~, the effect is obviously important.

We have decided not to include collisional rates to $n=5$ from the
unpublished results of SB93 as they are not converged. However, we
have identified the lines with $j/j_{4471} > 10^{-2}$ for which our
predictions would change significantly, by assuming that the collision
strength to each $n=5$ level is 50\% of that to the corresponding
$n=4$ level. For low densities ($n_{e}=10^{2}~cm^{-3}$), the intensity
of all but one emission line with $n_{u}=5$ increase by less than 3\%;
lines with $n_{u} \leq 4$ increase by less than 0.4\%.

For a density of $n_{e}=10^{6}~cm^{-3}$ and $t=2.0$, lines with
$n_{u}=5$ increase by up to 70\% from the tabulated predictions,
except for 4122\AA~ which is enhanced by a factor of 2.5. Several
bright lines are also enhanced by a few percent from the values in
Table 7 : 6678 \AA (1.0\%), 5876 \AA (4.8\%), 4471 (3.9\%). Thus,
unless one is interested in lines with $n_{u}=5$, or wishes to obtain
accuracies of better than 5\% for bright lines in nebulae with
$n_{e}>10^{6}~cm^{-3}$, the calculations presented here should be
sufficiently accurate.

Which collisional rates matter? We explored the effect of turning off
all collisional rates except for those with $\Delta n =0$ and
transitions from $2^{1}~S$ and $2^{3}~S$. We found a difference of at
most 0.5\% for lines with $j/j_{4471} > 10^{-2}$ over the temperature
range $t=0.5-2$ and density range $n_{e}=10^{0}-10^{8}~cm^{-3}$.  Thus
for this range of temperature and density, all other collisional
transitions may be safely ignored.

\subsection{Primordial helium abundance and uncertainties}

The calculations presented here provide a grid of emission
coefficients as a function of density and
temperature.  Assuming that optical depth effects are negligible,
ionization correction factors are accounted for, and given a temperature
and density from other spectral diagnostics, these
calculations may be combined with observed line fluxes to estimate 
the nebular helium abundance of an object. The fitting functions given in Table 5-7
should be sufficiently accurate for most applications, however for 
cosmological studies, we present even more accurate fits here together
with a detailed treatment of the uncertainties. 

In order to simplify our fitting function, we consider a restricted
range of temperatures and densities appropriate for low metallicity
extragalactic H II regions: $2 > t > 1.2$ and $300> n_{e}>1~{\rm
cm^{-3}}$.  We also consider the brightest optical lines which are not
likely to be severely affected by collisional or radiative transfer
effects: for singly ionized helium these are 4471 \AA, 5876 \AA, 6678 \AA, and
for doubly ionized helium, 4686 \AA. 

If the ratio of the observed helium line intensity to the $H \beta$
intensity is given by $r_{line}=I_{line}/I_{H \beta}$, then the helium
abundance by number using a given line is

\begin{equation}
y_{line}=r_{line} f_{line}(n_{e},t) ~,
\end{equation}

where the abundance by number can be converted to helium mass fraction
using the relationship of Pagel et al. (1992),

\begin{equation}
Y=\frac{4y[1-20(O/H)]}{1+4y}~,
\end{equation}

and $y=ICF \times (y^{+}+y^{++})$ is the ionization correction factor
(see Pagel et al 1992)
times the sum of $y^{+}$, the singly ionized helium fraction determined
using the He I lines, and $y^{++}$, the doubly ionized helium fraction
determined using He II 4686.

The function $f(n_{e},T)$ comes from combining our calculations with
the $H \beta$ emissivities of Storey \& Hummer (1995). We have
determined a fitting function for $f(n_{e},T)$ for the He I and He II
emission lines of the form $f=At^{B}$, where $B=B_{0}+B_{1}n_{e}$. The
parameters and goodness of fit statistics are given in Table 9. The
fitting function is accurate to within 1\% over the entire
temperature/density interval.

As emphasized by Olive, Skillman, \& Steigman (1997a,b), if the 
calculated helium emissivities are used to derive accurate helium
abundances, it is important to make detailed estimates for systematic
sources of uncertainty. Therefore, we need to assess the sources of 
error in the helium fraction

\begin{equation}
\sigma_{y}=\sqrt{\sigma_{f}^{2}r^{2}+\sigma_{r}^{2}f^{2}} 
\end{equation}

where $\sigma_{r}$ is the observational uncertainty.
The analysis uncertainty, $\sigma_{f}$, arises from
three sources which we add in quadrature: (1) uncertainties introduced
by the fact that we use a fitting function rather than the exact
tabulated emissivities, (2) uncertainties due to uncertainties in the 
atomic data,  (3) uncertainties derived from uncertainties
in the input density and temperature.  This can be written as 

\begin{equation}
\sigma_{f}^{2}=\sigma_{fit}^{2}+\sigma_{atomic}^{2}+
\sigma_{n}^2 \left(\frac{\partial f}{\partial n_{e}}\right)+
\sigma_{t}^2 \left(\frac{\partial f}{\partial t}\right) ~,
\end{equation}

where $\sigma_{n}$ and $\sigma_{t}$ are the uncertainties in the
electron density and normalized temperature, and the partial
derivatives are $\partial f/\partial t=ABt^{B-1}$ and $\partial
f/\partial n_{e}=A B_{1}(\ln{t})t^{B}$.  In the cases where there is a
significant slope with density and large uncertainties on $n_{e}$, it
is recommended that the uncertainty be determined by calculating the
value of $f$ over a range of densities, rather than using $\partial
f/\partial n$.

Strictly speaking, use of a fitting function introduces a systematic
rather than random uncertainty into our estimate.  Table 9
both the maximum absolute value of deviation over the fitting region
as well as the standard deviation so that one can limit the systematic
uncertainty.

Characterizing the uncertainties in all the atomic data that are
incorporated in this calculation is a difficult task. This is
principally because errors in the atomic data will almost certainly be
systematic rather than random, since the majority of data come from
numerical calculations which will have built-in approximations and
assumptions. As a result, calculated atomic data are rarely
reported with uncertainty estimates. Therefore, attempts to calculate
the uncertainty for an emission line intensity are somewhat subjective
and the error estimates given here should be considered illustrative
rather than definitive. Nevertheless, given the importance of
identifying and characterizing all the sources of uncertainty in
determination of primordial helium abundances, we have made such an attempt. 

For the restricted range of temperature and density in \S 3.2, we did a
set of calculations where for each run we randomly adjust each atomic
parameter each time, using the uncertainty estimates given in the
appendix.  The revised atomic parameter is assumed to be
Gaussian distributed about the given value with a standard deviation
given by the percentage confidence levels given in the previous
sections. After 1000 runs, we find that the mean line fluxes and
standard deviations are sufficiently converged. There is no
significant skewness in the resulting distribution.  We then average
the fractional standard deviation over the range of density and
temperature considered.  For the lines of interest, we find
$\sigma_{atomic}/f$ is in the range 1.3-1.5\%. The
greatest source of uncertainty at the current time is the
recombination rates to individual levels. These rates consist of a
patchwork of different approximations for different levels and
temperatures. A coherent treatment of these rates should be quite
valuable.

To illustrate the use of these uncertainties, we consider the case of
the low metallicity galaxy I Zw 18 using the data of Izotov \& Thuan
(1998). For the SE knot, they have determined $r_{6678}=0.0267 \pm 0.0010$ and 
$r_{4686}=0.0076 \pm 0.0009$. The [O III] determined temperature is 
$t=1.88 \pm 0.04$ and the density from [S II] is $n_{e}=10~cm^{-3}$.
Using the fitting function data from Table 9, we find $y^{+}=y_{6678}=0.0807$
$y^{++}=y_{4686}=6.80 \times 10^{-4}$. This agrees with the estimates of
Izotov \& Thuan (1998), since they also used the atomic data of S96,
and the measured densities are sufficiently low that collisional corrections
are unnecessary.

If we assume that the uncertainty in density is $\sigma_{n}=10
~cm^{-3}$, then the contributions to the flux ratio uncertainty
$\sigma_{f}^{2}$ are $\sigma_{fit}=3.05 \times 10^{-3}$,
$\sigma_{t}(\partial f/\partial t)=0.016$, $\sigma_{n}(\partial
f/\partial n)=3.82 \times 10^{-3}$, and $\sigma_{atomic}=0.039$, so
that it is the atomic data uncertainty that dominates this part of the
error budget, followed by the temperature. The uncertainty in $y$
arising from the analysis is therefore $\sigma_{f}=0.042$. In this
case, it is the measurement uncertainties that dominate the uncertainty in
$y$, since $\sigma_{f}r<<\sigma_{r}f$. For both source of uncertainty
to be equal, $\sigma_{f}$ would have to be 0.113 rather than the
calculated 0.042. 

In this example, a density uncertainty of $10~cm^{-3}$ is
unrealistically small. If the density uncertainties are
sufficiently large, it is necessary to use calculate the value of $f$
over the possible density range, rather than using
$\sigma_{n}(\partial f/\partial n)$. Using a density range of
$100 > n > 0 ~cm^{-3}$, we find that $f$ varies by $\Delta f=0.034$,
a significantly larger uncertainty than $\sigma_{n}(\partial
f/\partial n)=3.82 \times 10^{-3}$.

Frequently helium abundance are calculated by weighting the results from different lines by the respective uncertainties. This gives the greatest weight
to the brightest line measured, usually 5876 \AA. In the future, the
analysis uncertainties characterized here should also be included when 
doing this final weighting.

\section{Comparison to data and future work}

An ideal recombination model should be able to match all of the
observed helium emission line intensities simultaneously. Any
differences between the predictions and the observed line fluxes will
indicate either errors in the calculations or incompleteness in the
assumptions of our model.  In Figure 2, we plot the ratio of observed
helium emission line intensities to predictions for three nebulae: NGC
1976 in Orion (Osterbrock et al. 1992), NGC 6572, and IC 4997 (Hyung,
Aller \& Feibelman 1994a,b).  The predicted emisison lines are
calculated using $(n_{e},T_{e})=$ $(4000~cm^{-3},9000~K)$ for NGC
1976, $(10^{4}~cm^{-3},11000~K)$ for NGC 6572, and
$(10^{6}~cm^{-3},10800~K)$ for IC 4997. Figures 2a and 2b show the
ratio of observed-to-predicted lines ratios before and after
correcting for collisional excitation. Note that the ratio of
observed-to-predicted flux for 4471 is one by definition.  As expected
the inclusion of collisional effects results in better agreement for
the $n ^{3}S-2 ^{3} P$ series of lines (7065 \AA, 4713 \AA, etc) which
are most sensitive to collision excitation out of $2 ^{3} S$. The
7065/4471 ratio decreases from four to 1.8 times the predicted
ratio. Of course, the third ratio in this series, 4121/4471 \AA, gets
slightly worse. Our collision corrected lines agree well with those 
estimated by Kingdon \& Ferland (1996) which can be obtained by dividing
the first column by the second column tabulated for each object in their
Table 1.

After the collisional corrections have been applied, the most
discrepant bright optical lines are from the dense planetary nebula IC
4997: the ratios 7281/4471 and 6678/4471 are both less than
predicted. The origin of this discrepancy is not clear, but since the
discrepancy is monotonic with wavelength, Kingdon \& Ferland
(1996) have suggested that it is due to internal extinction associated
with an unusually high gas-to-dust ratio in this object.

Unfortunately, the observations to which we are comparing our
calculations had no error estimates for line fluxes; in order to
weight the significance of individual lines, we vary the point size
according to the measured brightness of the lines. Figure 2 shows that
the the infrared transitions to $n=3$ show roughly a factor of three
scatter evenly distributed about the predicted value. This scatter may
be due to measurement error, since these lines are faint and suffer
from uncertain telluric absorption. Without error estimates though, it
is hard to be sure that this is the entire explanation.

Finally, for the UV lines shortward of 4000 \AA, it appears that the
UV lines are systematically high. In particular, there appears to be
some curvature of the ratio upward as the wavelength of the line
decreases from 3614 \AA~ to 3354 \AA.  It seems likely that this is
also due to radiative transfer effects. Using HST spectra, Martin et al
(1996) have seen a similar effect, rule out extinction as a
possibility and suggest that radiative transfer effects are
present.

Over the last four years, high resolution
optical spectra, containing several helium lines each, and estimates
of physical conditions have been obtained for over a dozen planetary
nebulae by Hyung, Aller, \& Feibelman (see Hyung \& Aller 1997 and
references therein). We only compare to two of those nebulae here, but
clearly there is now a large data base with which to compare our
helium emissivity calculations. Such a data base will be be extremely
useful to see if deviations from our predictions correlate with
physical properties of the nebulae. A more detailed comparison of
our calculations to this expanded  database will be given in future
work.

What can be done to further improve the reliability of 
pedictions for helium emission lines? There are several
areas where the atomic model needs improvement or verification. First,
the recombination rates used are a patchwork of rates from different
sources, and are probably the single largest source of uncertainties
in the resulting emisison line intensities. Second, as discussed by
S96, strong singlet-triplet mixing can occur for high $l$
states. And finally, an independent evaluation of the collisional
rates for high $n$ states would be valuable, particularly as we
found discrepancies as high as a factor of seven in collisional 
rates using the Seaton (1962) impact parameter method and the {\bf R}-matrix
calculations of SB93.

Finally, in order to compare with observational measurements of helium
line intensties it is necessary to include radiative transfer effects
to determine what effects this has on the resulting line ratios.  All of
our results presented here are in the limit of Case B, in which all 
permitted transitions to $1 ^{1}S$ are thought to be immediately reabsorbed,
and all other transitions are optically thin.
Both nebulae with sufficiently low optical depth in transitions to $1^{1}S$ 
and nebulae with sufficiently high optical depth in $2 ^{3}S$ will depart
from this assumption. The
standard reference for radiative transfer are those of Robbins (1968)
and Robbins \& Bernat (1973). Recent examinations of this issue are
given by AN89, and also by Proga, Mikolajewska, \& Kenyon (1994) and
Sasselov \& Goldwirth (1995) who took their data from AN89. Given the
improvements in the atomic data afforded by the re-examination of
A-values (Kono \& Hattori 1984), the recombination rates (S96), and
collisional rates (SB93), a re-examination of radiative transfer
issues should be very useful. This will be the subject of future work.

ACKNOWLEDGEMENTS: We would like to acknowledge useful conversations
with Phil Stancil concerning the effect of charge transfer on
affecting the population of the $2 ^{3}S$ level, and Jim Kingdon and
Hagai Netzer for supplying codes, advice, and calculations, as well
as for numerous useful comments on this manuscript. 
We are also grateful for support from the Minnesota Supercomputer Institute
and NASA LTSARP grant no NAGW-3189.

\appendix
\centerline{ {\bf APPENDIX} }
\section{Atomic data}

The reliability of the calculations presented here is determined by
the accuracy of the input atomic data. In Tables 1-4, we give
a small subset of the atomic data for future researchers to compare with.
Here, we summarize the techniques and references for the rest of
the input data. 

\subsection{Energy levels ($E$)} 

 $l \leq 2$: When available, the energy levels from the compilation of
Martin (1973) were used. For levels above those given by Martin, $n_{top}$,
 energy levels were calculated
by extrapolating the quantum defect of these lower levels, 
$d=n-n^{*}$, where $n^{*}$ is the effective quantum number. The last
value of $n$ for which a term value is given, and the adopted quantum
defects are given in Table 10.

$l \geq 3$: Hydrogenic energy levels were assumed, i.e. $E=-E_{o}/n^{2}$, 
where $E_{o}=13.60577~eV$.

All energy levels should be accurate to within 0.5\%, and are given
in Table 1. Wavelengths are calculated from these energy levels. In
the tables, we use these vacuum wavelengths. In the text, we refer
to transitions using the standard wavelengths in air.

\subsection{Spontaneous radiative transitions ($A$)}

$n_{u} \leq 9$: Using values of Kono \& Hattori (1984).

$n_{u} \geq 10$ and $n_{l} \leq 6$: Using Coulomb approximation (see Smits
1991a), but each series of transitions to $n_{l}$ has been scaled to
match the Kono \& Hattori (1984) values for transitions from $n_{u}=9$
to $n_{l}$.

$n_{u} \geq 10$ and $n_{l} \geq 7$ : Using Coulomb approximations as in
Smits (1991a).

{\it Non-permitted transitions}: Rates for the transition from $2^{3}S$ to 
$1^{1}S$ is from Hata \& Grant (1981), $2^{3}P$ to $1^{1}S$ from Lin,
Johnson, \& Dalgarno (1977) and the two-photon transition 
from $2^{1}S$ to $1^{1}S$ is from Drake (1979). The remaining forbidden transitions 
among the $n=2$ levels are taken from Mendoza (1983).

Nearly all previous calculations of helium emission lines before
Smits used A values taken from the compilation of Wiese, Smith, \& Glennon
(1966). Many of these values derived from the calculations of Dalgarno \&
Kingston (1958). There are numerous cases, such as the 3187 \AA~line, in
which the A-values differ by more than 10\%. Therefore, if accuracies of
a few percent or better are desired, these new calculations should be used.
Kono \& Hattori (1984) estimate all values are accurate to within 1\%,
and many to better then 0.1\%. Here we adopt an uncertainty of 1\%
for all lines.

\subsection{Radiative recombination rates ($\alpha$)}

$^{1}$S, $^{3}$S n $\leq$ 10; $^{1}$P, $^{1}$D, $^{3}$D n $\leq$ 9;
$^{3}$P $n = 2,3$: Recombination rates obtained by detailed balance from
He I photoionization cross sections of Fernley et al. (1987) using
integration scheme outlined by Burgess (1965) and Brocklehurst (1971).
Tabulated values of the photoionization cross-sections $\sigma$ and
energies $\omega$ were obtained from TOPbase (Cunto et al. 1993), and
the necessary grid of points needed for the integration was obtained
by linearly interpolating $\log \sigma$ vs $\log \omega$.  Method described
in S96.

$^{3}$P $n = 4$ to 12, $^{1}$P $n = 10$, 11, 12: Quantum defect method
was used: for $T < 2500~{\rm K}$ algorithm of Burgess \& Seaton (1960a) and
for $T \geq 2500~{\rm K}$ algorithm of Peach (1967)

Remaining levels: Scaled hydrogenic rates used for large $n$.  Scaling
factor was determined from ratio of TOPbase to hydrogen recombination
rate at $n = 10$ for S levels and $n = 9$ for D levels.  For the P
series with $n > 12$ scaled hydrogenic rates were used, scaling factor
taken from quantum defect method results at $n = 12$.  Hydrogenic
rates are described in S91.

There have been several works that examine the \underline{total}
photoionization cross section as a function of energy. Opacity Project
calculations (Fernley et al 1987; Cunto et al 1993) used to derive the
photoionization cross section have been compared with observational
data from threshhold to 120 eV showing differences of less than 1\%
(Samson et al. 1994). Samson et al. then critically evaluate the
available literature to extend the cross sections to 8 keV.  Yan et al
(1998) have performed theoretical calculations at these higher
energies, and suggested revision in the Samson et al cross section.

In comparing calculations of cross-section to \underline{individual}
levels at the threshhold energies, Hummer \& Storey (1998) found that
the Fernley et al. (1987) calculations agreed to within 1\% with
results expected using the oscillator strengths of Drake
(1996). However, much larger differences than this were found for the
$1 ^{1}S$ series, which motivated Hummer \& Storey (1998) to perform
ab initio calculations so as to provide more reliable cross-sections
for the individual levels. Although the resulting individual
recombination rates are not yet available, we can compare the total
recombination rate here with the values from Hummer \& Storey
(1998). At $T=10^{4}~K$, their value for $\alpha_{B}^{1}$ and 
$\alpha_{B}^{3}$ are 1.3\% and 1.1\% higher than our values.

In our calculations, we esitmate that uncertainties in the individual
rates are probably of order 2\%. This uncertainly can eventually
be reduced by using the new Hummer \& Storey rates which would eliminate
the heterogeneous treatment of individual rates. 

\subsection{Collision strengths ($\Omega$)}

$\Delta S \neq 0$ and $n_{u} \geq 5$: No collisional coupling 
between the singlet and triplet state is considered for $n_{u} \geq 5$.

$\Delta S \neq 0$ and $n_{u} \leq 4$: Coupling between all triplet and
singlet levels for $n_{u} \leq 4$ due to electron collisions only is
taken from SB93, by linearly interpolating
$\Omega$ from their Table II. For temperatures below the 
minimum calculated by SB93, $\Omega=0$; for temperatures above the 
maximum temperature, $T_{max}$, $\Omega=\Omega(T_{max})$. This is
done to minimize the effects of uncertain extrapolation.

$\Delta n \neq 0$, and $n_{u} \geq 5$: Only electronic transitions are
considered.  Hydrogenic collision rates are used for $l \geq 3$;
impact-parameter treatment of Seaton (1962) is used for $l \leq 2$.

$\Delta n \neq 0$ and $n_{u} \leq 4$: Only electronic transitions 
are included and are taken from SB93. 

$\Delta n = 0$ and $n \geq 2$: Transitions involving electrons,
protons, and ionized helium atoms are included, assuming
$n_{e}=n_{p}=10n_{He^{+}}$. Although this assumption technically
makes the emission lines predictions dependent on chemical abundance
and ionization state, the dependence is normally so weak as to be
negligible. For a given $n$, the low $l$ values are calculated using
the impact parameter method of Seaton (1962); when a value of $l$ is
reached such that the computationally faster method of
Pengelly \& Seaton (1964) matches the Seaton (1962) method within 6\%, the second
method is used for all higher $l$ states. We do not use SB93 rates
since only a few of these rates at low temperature are judged by them
to be converged.

$\Delta n = 0$ and $n=2$: We use the electron collision rates of
SB93, and the method of Seaton (1962) for the proton and helium
atom collisions. The energy difference is large enough that electron
collisions completely dominate the total.

Table 4 contains the collision strengths used for a range of
temperature together with power-law fits. These collision strengths
may be converted to collisional rates using the standard formula in
Osterbrock (1989,p. 55). Note, however, that many of
the rates of SB93 are not monotonic with
temperature, since they integrate a Maxwellian distribution of
electron velocities over a cross-section with resonances. As a result, any
attempt to fit these rates with simple power laws (as in Clegg 1987 or
Kingdon \& Ferland 1996) will necessarily result in a loss of
accuracy. Whether this loss of accuracy is significant depends upon
the relative importance of collisional processes in determining the
line of interest. It is also worth noting that in the process of
comparing $\Delta n \neq 0$ collisional rates using Seaton impact parameter
method (1962) to the more
accurate values of SB93, we found differences of up to a factor
of seven in the electron impact collision strengths. However, there were
not enough transitions in common for the two data sets to search for
systematic effects with quantum number. For the collision strengths of
SB93, we assume an uncertainty of 2\%. For all other collision strengths,
we assume a very optimistic 20\%.

\subsection{Collisional ionization ($C^{ion}$)}  Following AN89, we use the collisional ionization rate of Taylor, Kingston \& Bell
(1979) for the $2^{3}S$ level, and scaled hydrogenic rates of Vriens
\& Smeet (1980) for all other rates. Collisional ionization becomes
most significant at high densities for high $n$ levels; however, the
calculations of S96 which are used to calculate the effect of upper
levels on the recombination cascade do not currently include
collisional ionization. Fits to the collisional rates of the
form $C^{ion}=a_{C} t^{b_{C}} e^{C_{C}/t}$ are given in Table 2.

\subsection{Three-body recombination ($\alpha^{tbr}$)} Three-body recombination
is calculated using detailed balance from the above rates, fit as
$\alpha^{tbr}=\alpha_{4}^{tbr} t^{b_{\alpha^{tbr}}}$ and given in
Table 2.

\subsection{Other possible rates}
There are several other possible physical effects that could be 
considered here, but are not included in this work. 

{\it Charge transfer}: Clegg (1987) has discussed the possibility that
charge transfer could be important in affecting the $2^{3}S$ population.
For reactions of the form $X~+He(2^{3}S) \rightarrow$ products, he is 
able to rule out the importance of this transition if X is a metal, and
standard nebular abundances are assumed. (For a case in which this 
assumption breaks down see models of H-poor supernova ejecta by Swartz 1994.)
Using the rates for proton collisions of Janev et al. (1987), Baldwin et al. 
(1991) also shows that proton collision exchange is also negligible.

{\it Photoionization from metastable levels}: Clegg \& Harrington
(1989) have also considered the effects of photoionization from the
metastable $2^{3}S$ level. This effect can be considerable in optically-thick 
planetary nebulae, but is found to be negligible for optically-thin 
planetaries and low density H II regions. We neglect these effects here
as it requires detailed modelling of the nebular spectrum.

{\it Dielectronic recombination}: At sufficiently high temperatures,
$T >50,000~K$, dielectronic recombination starts to become important
relative to radiative recombination (Burgess 1964). The
excitation of the helium atom to an autoionizing state will change the
energy level structure and the resulting cascade. This complication
is not addressed in this paper, but is considered in Bhatia \&
Underhill (1986).

{\it Radiative absorption}: A final effect which has been to found to
be important (Robbins 1968, Robbins \& Bernat 1973) is the
absorption of line photons from the ground state $1^{1}S$ and the
metastable levels $2^{1}S$ and $2^{3}S$. In this work, we assume that
all line photons associated with decays to $1^{1}S$ are reabsorbed on
the spot (Case B). Treatment of absorption of line photons from the 
$2^{1}S$ and $2^{3}S$ levels is deferred to later work.

\vfill\eject

\begin{figure}
\plotone{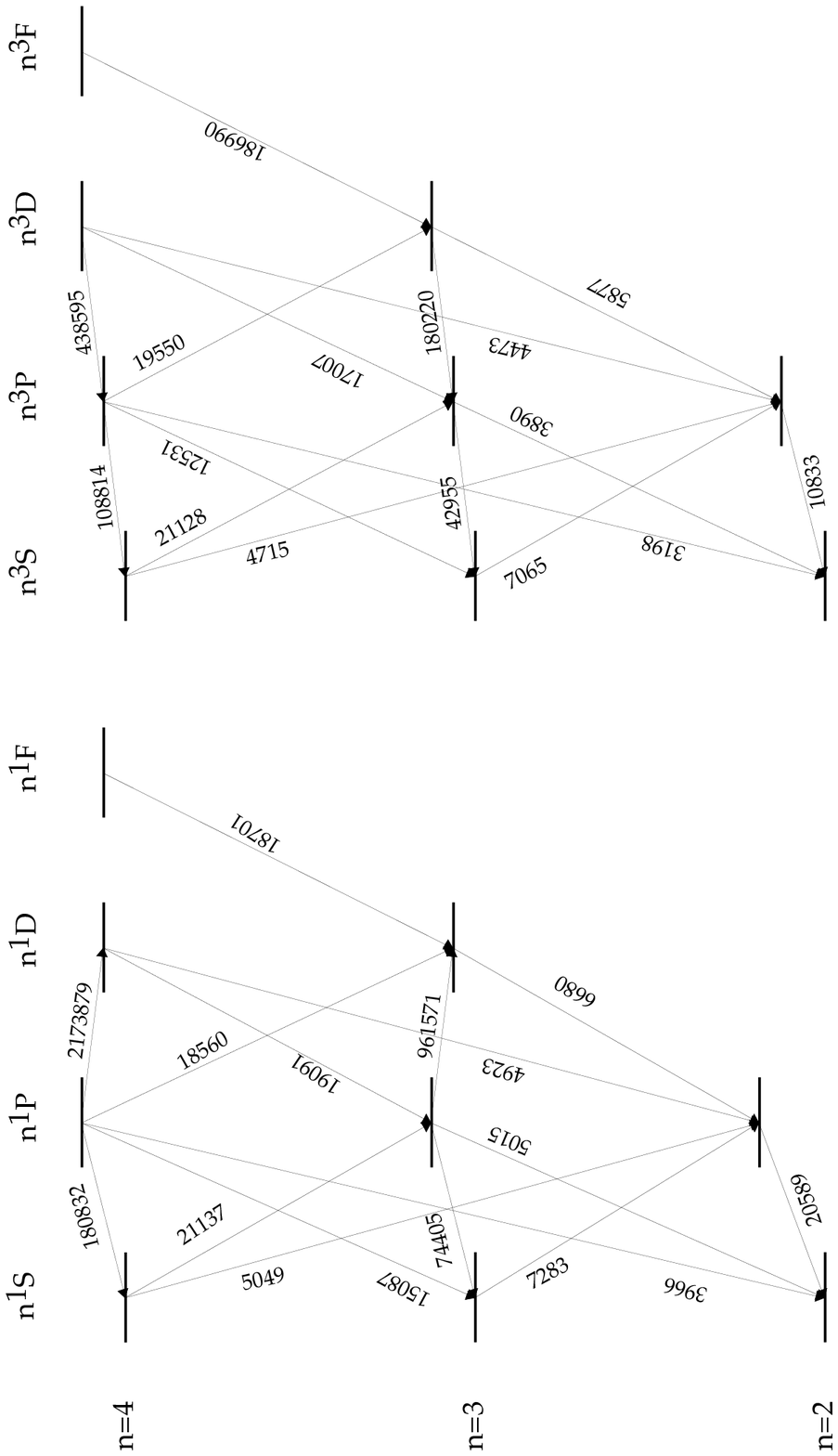}
\caption{Grotrian diagram for He I singlet and triplet ladders for 
levels up to n=4. Wavelengths are vacuum wavelenghts as calculated from
the energy levels in Table 1. ``Families'' of lines, such as the 
$n ^{3}P-2 ^{3}S$ series, can be found collected together in Table
4.}
\end{figure}\label{fig-grotrian}

\begin{figure}
\figurenum{2a}
\plotone{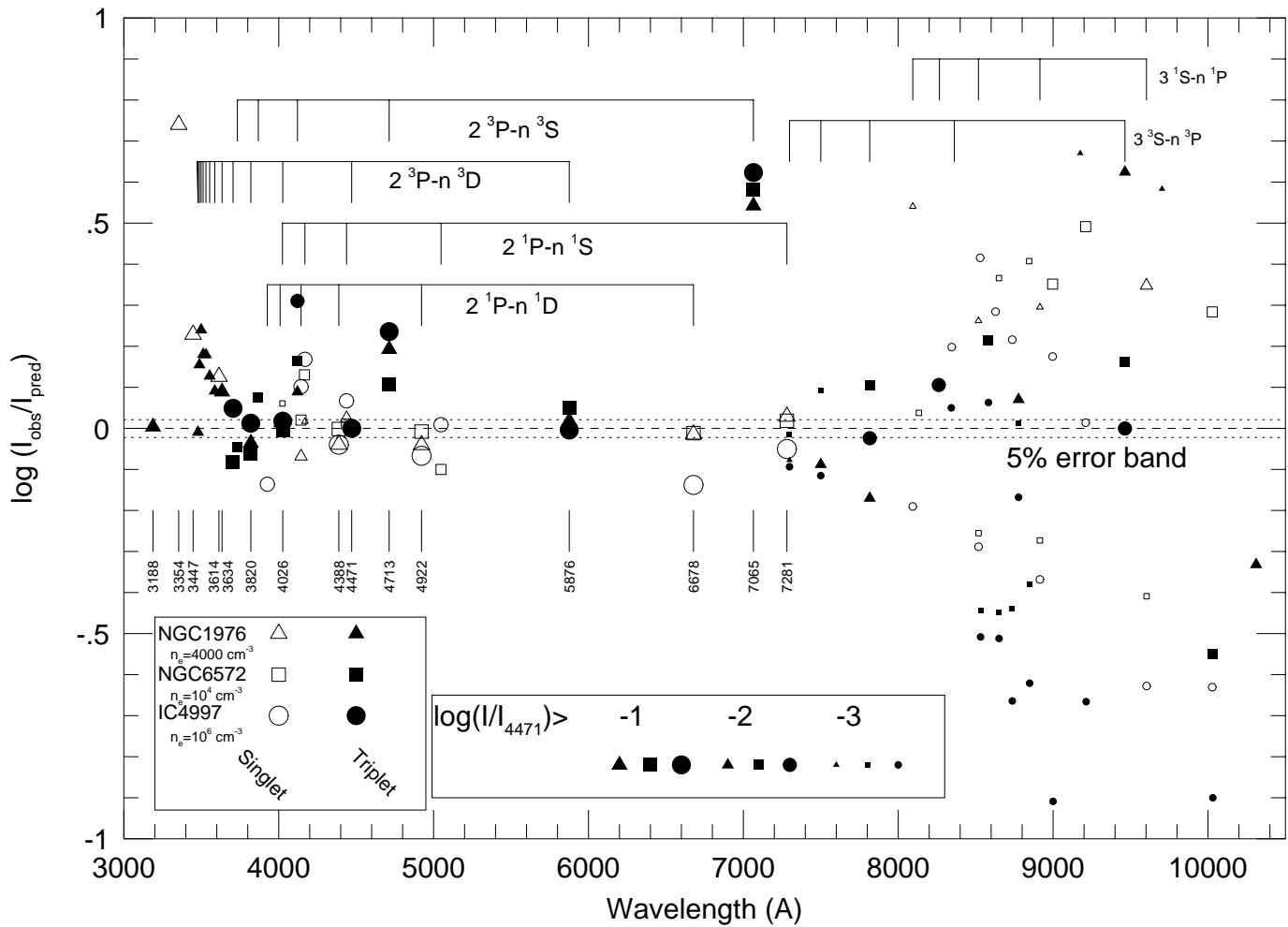}
\caption{Comparison between theoretical emissivities (without collisional
corrections) and observed
helium emission line intensities for three nebula: Orion (NGC 1976)
with $n_{e}=4000~cm^{-3}$ and $T_{e}=9,000$ K, and the planetaries NGC
6572 ($n_{e}=10^{4}~cm^{-3}$,$T_{e}=11,000$ K) and IC4997
($n_{e}=10^{6}~cm^{-3}$,$T_{e}=10,800$ K). Singlet and triplet lines
are shown with open and filled symbols respectively. All lines have
been normalized to 4471 \AA, and where the symbol size indicates the
brightness of the lines.  Selected ``families'' of lines are noted at
the top of the plot, while the observed wavelengths of selected lines
are noted at the bottom. Ideally, all the points should lie within the
5\% error band. See text for further discussion.}
\end{figure}\label{fig-hei}

\begin{figure}
\figurenum{2b}
\plotone{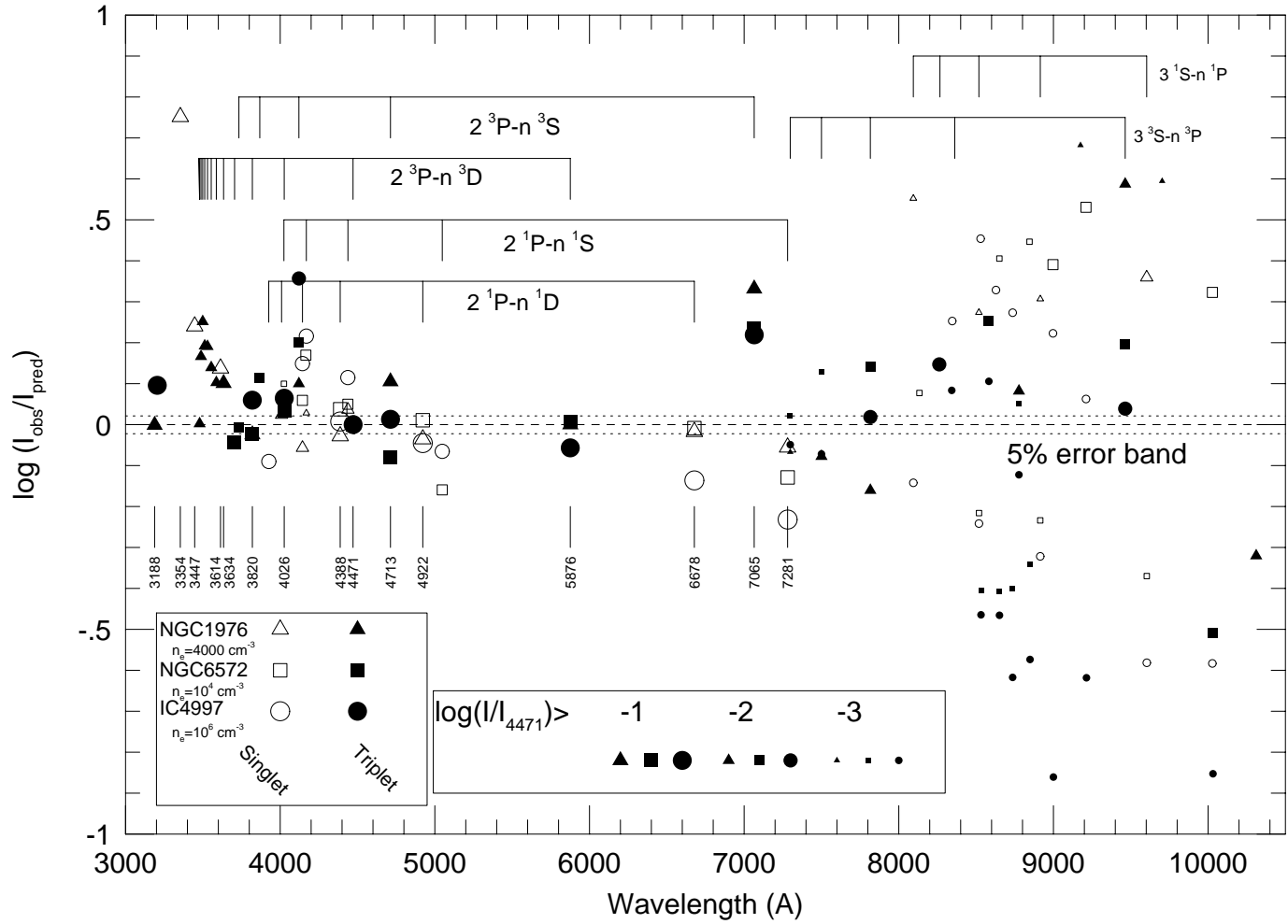}
\caption{Same as Figure 2a, but after collision excitation from 
$2 ^{3}S$ and $2 ^{1}S$ have been included}
\end{figure} 



\vfill\eject



\makeatletter
\def\jnl@aj{AJ}
\ifx\revtex@jnl\jnl@aj\let\tablebreak=\nl\fi
\makeatother


\begin{planotable}{lrrrrcr}
\tablecaption{Recombination data for individual levels of He I}
\tablehead{
\colhead{n}           & \colhead{l}      &
\colhead{$g_{S}$}          & \colhead{$E_{nls}$}  &
\colhead{$\alpha_{4}$}          & \colhead{b$_{\alpha}$}    &
\colhead{e$_{\alpha}$}  \\
\colhead{}           & \colhead{}      &
\colhead{}          & \colhead{(eV)}  &
\colhead{(${\rm cm^{3}~s^{-1}}$)}          & \colhead{}    &
\colhead{}  
}

\startdata
  & $ \alpha^{tot}_{A} $ &  &  & $   4.27  \times 10^{ -13}   $ &  -0.678 & $ \pm 0.00 \% $ \nl
  & $ \alpha^{tot}_{B} $ &  &  & $   2.72  \times 10^{   -13} $ &  -0.789 & $ \pm 0.01 \% $  \nl
  & $ \alpha^{1}_{B} $ &  &  & $   6.23  \times 10^{   -14} $ &  -0.827 & $ \pm 0.02 \% $ \nl
  1 &  0 &  1 & -24.5876 & $   1.54 \times 10^{   -13} $ & -0.486 & $ \pm 0.27 \% $  \nl
  2 &  0 &  1 &  -3.9716 & $   5.55 \times 10^{   -15} $ & -0.451 & $ \pm 0.34 \% $  \nl
  2 &  1 &  1 &  -3.3694 & $   1.26 \times 10^{   -14} $ & -0.695 & $ \pm 3.36 \% $  \nl
  3 &  0 &  1 &  -1.6671 & $   1.62 \times 10^{   -15} $ & -0.444 & $ \pm 0.23 \% $  \nl
  3 &  1 &  1 &  -1.5005 & $   5.01 \times 10^{   -15} $ & -0.700 & $ \pm 3.34 \% $  \nl
  3 &  2 &  1 &  -1.5134 & $   4.31 \times 10^{   -15} $ & -0.872 & $ \pm 4.77 \% $  \nl
  4 &  0 &  1 &  -0.9139 & $   7.00 \times 10^{   -16} $ & -0.445 & $ \pm 0.08 \% $  \nl
  4 &  1 &  1 &  -0.8453 & $   2.43 \times 10^{   -15} $ & -0.708 & $ \pm 3.39 \% $  \nl
  4 &  2 &  1 &  -0.8510 & $   2.67 \times 10^{   -15} $ & -0.871 & $ \pm 4.80 \% $  \nl
  4 &  3 &  1 &  -0.8504 & $   1.38 \times 10^{   -15} $ & -1.046 & $ \pm 5.25 \% $  \nl
  5 &  0 &  1 &  -0.5762 & $   3.66 \times 10^{   -16} $ & -0.454 & $ \pm 0.23 \% $  \nl
  5 &  1 &  1 &  -0.5416 & $   1.35 \times 10^{   -15} $ & -0.718 & $ \pm 3.42 \% $  \nl
  5 &  2 &  1 &  -0.5447 & $   1.60 \times 10^{   -15} $ & -0.873 & $ \pm 4.81 \% $  \nl
  5 &  3 &  1 &  -0.5442 & $   1.22 \times 10^{   -15} $ & -1.048 & $ \pm 5.25 \% $  \nl
  5 &  4 &  1 &  -0.5442 & $   4.46 \times 10^{   -16} $ & -1.183 & $ \pm 4.80 \% $  \nl
  & $  \alpha^{3}_{B} $ &  &  & $   2.10  \times 10^{   -13} $ &  -0.778 & $ \pm 0.01 \% $ \nl
  2 &  0 &  3 &  -4.7679 & $   1.49 \times 10^{   -14} $ & -0.381 & $ \pm 1.41 \% $  \nl
  2 &  1 &  3 &  -3.6233 & $   5.61 \times 10^{   -14} $ & -0.639 & $ \pm 2.45 \% $  \nl
  3 &  0 &  3 &  -1.8690 & $   3.72 \times 10^{   -15} $ & -0.344 & $ \pm 1.38 \% $  \nl
  3 &  1 &  3 &  -1.5803 & $   1.95 \times 10^{   -14} $ & -0.632 & $ \pm 2.41 \% $  \nl
  3 &  2 &  3 &  -1.5137 & $   1.33 \times 10^{   -14} $ & -0.868 & $ \pm 4.72 \% $  \nl
  4 &  0 &  3 &  -0.9935 & $   1.50 \times 10^{   -15} $ & -0.328 & $ \pm 0.90 \% $  \nl
  4 &  1 &  3 &  -0.8796 & $   9.04 \times 10^{   -15} $ & -0.632 & $ \pm 2.47 \% $  \nl
  4 &  2 &  3 &  -0.8513 & $   8.29 \times 10^{   -15} $ & -0.867 & $ \pm 4.74 \% $  \nl
  4 &  3 &  3 &  -0.8504 & $   4.15 \times 10^{   -15} $ & -1.046 & $ \pm 5.25 \% $  \nl
  5 &  0 &  3 &  -0.6155 & $   7.51 \times 10^{   -16} $ & -0.327 & $ \pm 0.57 \% $  \nl
  5 &  1 &  3 &  -0.5592 & $   4.84 \times 10^{   -15} $ & -0.636 & $ \pm 2.47 \% $  \nl
  5 &  2 &  3 &  -0.5448 & $   4.99 \times 10^{   -15} $ & -0.870 & $ \pm 4.75 \% $  \nl
  5 &  3 &  3 &  -0.5442 & $   3.65 \times 10^{   -15} $ & -1.048 & $ \pm 5.25 \% $  \nl
  5 &  4 &  3 &  -0.5442 & $   1.34 \times 10^{   -15} $ & -1.183 & $ \pm 4.80 \% $  
\tablecomments{Radiative recombination rates and three-body recombination rate for individual levels can be determined by powerlaw fit
$\alpha(t_{4})=\alpha_{4}t_{4}^b$.  The maximum error of these fits
over the temperature range $T=5000-20000~K$ is $\pm e_{\alpha}$.
Calculations requiring higher accuracy or temperatures outside this
range should contact authors for a grid of rates.}
\end{planotable}



\makeatletter
\def\jnl@aj{AJ}
\ifx\revtex@jnl\jnl@aj\let\tablebreak=\nl\fi
\makeatother


\begin{planotable}{lrrrrrrrrr}
\tablecaption{Three body recombination and collisional ionization rates for He I}
\tablehead{
\colhead{n}           & \colhead{l}      &
\colhead{$g_{S}$}      & \colhead{$\alpha^{tbr}_{4}$}  &
\colhead{b$_{\alpha^{tbr}}$}  & \colhead{e$_{\alpha^{tbr}}$} & 
\colhead{$a_{C^{ion}}$}  & \colhead{$b_{C^{ion}}$} &
\colhead{$c_{C^{ion}}$} & \colhead{$e_{C^{ion}}$} \\
\colhead{}           & \colhead{}      &
\colhead{}          &  \colhead{(${\rm cm^{6}~s^{-1}}$)}  &
\colhead{}  & \colhead{} & 
\colhead{(${\rm cm^{3}~s^{-1}}$)} & \colhead{} &
\colhead{} & \colhead{}}
\startdata
  1 &  0 &  1 & $   3.18 \times 10^{   -31} $ & -0.905 & $ \pm 2.45 \% $ &  $   3.36 \times 10^{    -9} $ &  0.499 & -28.625 & $ \pm 0.18\% $ \nl
  2 &  0 &  1 & $   1.21 \times 10^{   -29} $ & -1.117 & $ \pm 3.36 \% $ &  $   1.33 \times 10^{    -7} $ &  0.253 &  -4.733 & $ \pm 0.21\% $ \nl
  2 &  1 &  1 & $   4.96 \times 10^{   -29} $ & -1.138 & $ \pm 3.42 \% $ &  $   1.81 \times 10^{    -7} $ &  0.229 &  -4.037 & $ \pm 0.22\% $ \nl
  3 &  0 &  1 & $   5.94 \times 10^{   -29} $ & -1.236 & $ \pm 3.70 \% $ &  $   6.58 \times 10^{    -7} $ &  0.120 &  -2.072 & $ \pm 0.23\% $ \nl
  3 &  1 &  1 & $   2.14 \times 10^{   -28} $ & -1.252 & $ \pm 3.75 \% $ &  $   7.93 \times 10^{    -7} $ &  0.103 &  -1.880 & $ \pm 0.24\% $ \nl
  3 &  2 &  1 & $   3.52 \times 10^{   -28} $ & -1.251 & $ \pm 3.74 \% $ &  $   7.81 \times 10^{    -7} $ &  0.104 &  -1.895 & $ \pm 0.24\% $ \nl
  4 &  0 &  1 & $   1.67 \times 10^{   -28} $ & -1.326 & $ \pm 3.95 \% $ &  $   1.87 \times 10^{    -6} $ &  0.021 &  -1.207 & $ \pm 0.25\% $ \nl
  4 &  1 &  1 & $   5.72 \times 10^{   -28} $ & -1.338 & $ \pm 3.99 \% $ &  $   2.13 \times 10^{    -6} $ &  0.007 &  -1.128 & $ \pm 0.25\% $ \nl
  4 &  2 &  1 & $   9.42 \times 10^{   -28} $ & -1.337 & $ \pm 3.98 \% $ &  $   2.11 \times 10^{    -6} $ &  0.008 &  -1.135 & $ \pm 0.25\% $ \nl
  4 &  3 &  1 & $   1.32 \times 10^{   -27} $ & -1.338 & $ \pm 3.98 \% $ &  $   2.11 \times 10^{    -6} $ &  0.008 &  -1.134 & $ \pm 0.25\% $ \nl
  5 &  0 &  1 & $   3.57 \times 10^{   -28} $ & -1.399 & $ \pm 4.13 \% $ &  $   4.02 \times 10^{    -6} $ & -0.059 &  -0.821 & $ \pm 0.25\% $ \nl
  5 &  1 &  1 & $   1.18 \times 10^{   -27} $ & -1.409 & $ \pm 4.15 \% $ &  $   4.44 \times 10^{    -6} $ & -0.069 &  -0.781 & $ \pm 0.25\% $ \nl
  5 &  2 &  1 & $   1.95 \times 10^{   -27} $ & -1.408 & $ \pm 4.14 \% $ &  $   4.40 \times 10^{    -6} $ & -0.068 &  -0.785 & $ \pm 0.25\% $ \nl
  5 &  3 &  1 & $   2.74 \times 10^{   -27} $ & -1.408 & $ \pm 4.15 \% $ &  $   4.40 \times 10^{    -6} $ & -0.068 &  -0.784 & $ \pm 0.25\% $ \nl
  5 &  4 &  1 & $   3.52 \times 10^{   -27} $ & -1.408 & $ \pm 4.15 \% $ &  $   4.40 \times 10^{    -6} $ & -0.068 &  -0.784 & $ \pm 0.25\% $ \nl
  2 &  0 &  3 & $   2.57 \times 10^{   -29} $ & -1.093 & $ \pm 3.29 \% $ &  $   9.36 \times 10^{    -8} $ &  0.280 &  -5.655 & $ \pm 0.21\% $ \nl
  2 &  1 &  3 & $   1.30 \times 10^{   -28} $ & -1.129 & $ \pm 3.40 \% $ &  $   1.58 \times 10^{    -7} $ &  0.240 &  -4.331 & $ \pm 0.21\% $ \nl
  3 &  0 &  3 & $   1.45 \times 10^{   -28} $ & -1.220 & $ \pm 3.66 \% $ &  $   5.36 \times 10^{    -7} $ &  0.138 &  -2.304 & $ \pm 0.23\% $ \nl
  3 &  1 &  3 & $   5.87 \times 10^{   -28} $ & -1.244 & $ \pm 3.73 \% $ &  $   7.23 \times 10^{    -7} $ &  0.111 &  -1.972 & $ \pm 0.23\% $ \nl
  3 &  2 &  3 & $   1.06 \times 10^{   -27} $ & -1.250 & $ \pm 3.74 \% $ &  $   7.81 \times 10^{    -7} $ &  0.104 &  -1.895 & $ \pm 0.24\% $ \nl
  4 &  0 &  3 & $   4.36 \times 10^{   -28} $ & -1.314 & $ \pm 3.92 \% $ &  $   1.62 \times 10^{    -6} $ &  0.035 &  -1.298 & $ \pm 0.25\% $ \nl
  4 &  1 &  3 & $   1.61 \times 10^{   -27} $ & -1.332 & $ \pm 3.97 \% $ &  $   2.00 \times 10^{    -6} $ &  0.014 &  -1.167 & $ \pm 0.25\% $ \nl
  4 &  2 &  3 & $   2.83 \times 10^{   -27} $ & -1.337 & $ \pm 3.98 \% $ &  $   2.11 \times 10^{    -6} $ &  0.008 &  -1.135 & $ \pm 0.25\% $ \nl
  4 &  3 &  3 & $   3.96 \times 10^{   -27} $ & -1.338 & $ \pm 3.98 \% $ &  $   2.11 \times 10^{    -6} $ &  0.008 &  -1.134 & $ \pm 0.25\% $ \nl
  5 &  0 &  3 & $   9.63 \times 10^{   -28} $ & -1.388 & $ \pm 4.11 \% $ &  $   3.61 \times 10^{    -6} $ & -0.047 &  -0.866 & $ \pm 0.25\% $ \nl
  5 &  1 &  3 & $   3.37 \times 10^{   -27} $ & -1.404 & $ \pm 4.14 \% $ &  $   4.21 \times 10^{    -6} $ & -0.064 &  -0.802 & $ \pm 0.25\% $ \nl
  5 &  2 &  3 & $   5.85 \times 10^{   -27} $ & -1.408 & $ \pm 4.14 \% $ &  $   4.39 \times 10^{    -6} $ & -0.068 &  -0.785 & $ \pm 0.25\% $ \nl
  5 &  3 &  3 & $   8.21 \times 10^{   -27} $ & -1.408 & $ \pm 4.15 \% $ &  $   4.40 \times 10^{    -6} $ & -0.068 &  -0.784 & $ \pm 0.25\% $ \nl
  5 &  4 &  3 & $   1.06 \times 10^{   -26} $ & -1.408 & $ \pm 4.15 \% $ &  $   4.40 \times 10^{    -6} $ & -0.068 &  -0.784 & $ \pm 0.25\% $ \nl
\end{planotable}




\makeatletter
\def\jnl@aj{AJ}
\ifx\revtex@jnl\jnl@aj\let\tablebreak=\nl\fi
\makeatother


\begin{planotable}{cccrrrrr}
\tablecaption{``Indirect''\tablenotemark{a} recombination rates for $n_{max}=5$(Case B) \tablenotemark{b}}
\tablehead{
\colhead{n}           & \colhead{l}      &
\colhead{$g_{S}$}          & \colhead{$\alpha^{\prime}_{4}$}  &
\colhead{$b_{\alpha}^{\prime}$}        &
\colhead{$e_{\alpha^{\prime}}(n_{e}=10^{2})$}~~~~\tablenotemark{c}  & 
\colhead{$c_{\alpha^{\prime}}$} &
\colhead{$e_{\alpha^{\prime}}(T_{e}=10^{4})$}~~~~\tablenotemark{d}  \\
\colhead{}           & \colhead{}      &
\colhead{}          & \colhead{(${\rm cm^{3} s^{-1}}$)}  &
\colhead{}        &
\colhead{}  & 
\colhead{} &
\colhead{} }
\startdata
  2 &  0 &  1 & $   2.25 \times 10^{   -15} $ & -0.766 & $ \pm 2.84 \% $ &  0.000 & $ \pm 0.11 \% $ \nl
  2 &  1 &  1 & $   3.11 \times 10^{   -15} $ & -0.879 & $ \pm 3.46 \% $ &  0.008 & $ \pm 3.49 \% $ \nl
  3 &  0 &  1 & $   6.09 \times 10^{   -16} $ & -0.782 & $ \pm 2.53 \% $ &  0.000 & $ \pm 0.00 \% $ \nl
  3 &  1 &  1 & $   1.30 \times 10^{   -15} $ & -0.876 & $ \pm 2.70 \% $ &  0.008 & $ \pm 3.15 \% $ \nl
  3 &  2 &  1 & $   2.64 \times 10^{   -15} $ & -1.114 & $ \pm 4.28 \% $ &  0.003 & $ \pm 1.08 \% $ \nl
  4 &  0 &  1 & $   2.44 \times 10^{   -16} $ & -0.783 & $ \pm 2.52 \% $ &  0.000 & $ \pm 0.04 \% $ \nl
  4 &  1 &  1 & $   6.50 \times 10^{   -16} $ & -0.854 & $ \pm 2.44 \% $ &  0.008 & $ \pm 3.06 \% $ \nl
  4 &  2 &  1 & $   1.64 \times 10^{   -15} $ & -1.106 & $ \pm 4.20 \% $ &  0.003 & $ \pm 1.08 \% $ \nl
  4 &  3 &  1 & $   2.31 \times 10^{   -15} $ & -1.271 & $ \pm 3.44 \% $ & -0.001 & $ \pm 0.15 \% $ \nl
  5 &  0 &  1 & $   1.16 \times 10^{   -16} $ & -0.786 & $ \pm 2.47 \% $ &  0.000 & $ \pm 0.11 \% $ \nl
  5 &  1 &  1 & $   3.82 \times 10^{   -16} $ & -0.832 & $ \pm 2.15 \% $ &  0.008 & $ \pm 2.93 \% $ \nl
  5 &  2 &  1 & $   9.86 \times 10^{   -16} $ & -1.096 & $ \pm 4.08 \% $ &  0.003 & $ \pm 1.09 \% $ \nl
  5 &  3 &  1 & $   1.94 \times 10^{   -15} $ & -1.267 & $ \pm 3.47 \% $ & -0.001 & $ \pm 0.13 \% $ \nl
  5 &  4 &  1 & $   2.80 \times 10^{   -15} $ & -1.375 & $ \pm 1.96 \% $ & -0.007 & $ \pm 0.78 \% $ \nl
  2 &  0 &  3 & $   7.07 \times 10^{   -15} $ & -0.675 & $ \pm 2.05 \% $ & -0.001 & $ \pm 0.20 \% $ \nl
  2 &  1 &  3 & $   1.01 \times 10^{   -14} $ & -0.901 & $ \pm 3.89 \% $ &  0.009 & $ \pm 4.16 \% $ \nl
  3 &  0 &  3 & $   1.51 \times 10^{   -15} $ & -0.696 & $ \pm 1.63 \% $ & -0.001 & $ \pm 0.34 \% $ \nl
  3 &  1 &  3 & $   3.63 \times 10^{   -15} $ & -0.873 & $ \pm 2.50 \% $ &  0.010 & $ \pm 3.89 \% $ \nl
  3 &  2 &  3 & $   8.19 \times 10^{   -15} $ & -1.097 & $ \pm 3.98 \% $ &  0.002 & $ \pm 1.01 \% $ \nl
  4 &  0 &  3 & $   5.16 \times 10^{   -16} $ & -0.701 & $ \pm 1.56 \% $ & -0.001 & $ \pm 0.46 \% $ \nl
  4 &  1 &  3 & $   1.62 \times 10^{   -15} $ & -0.845 & $ \pm 2.06 \% $ &  0.010 & $ \pm 3.88 \% $ \nl
  4 &  2 &  3 & $   5.19 \times 10^{   -15} $ & -1.078 & $ \pm 3.73 \% $ &  0.002 & $ \pm 0.98 \% $ \nl
  4 &  3 &  3 & $   6.94 \times 10^{   -15} $ & -1.271 & $ \pm 3.44 \% $ & -0.001 & $ \pm 0.17 \% $ \nl
  5 &  0 &  3 & $   2.09 \times 10^{   -16} $ & -0.711 & $ \pm 1.39 \% $ & -0.001 & $ \pm 0.69 \% $ \nl
  5 &  1 &  3 & $   8.63 \times 10^{   -16} $ & -0.813 & $ \pm 1.51 \% $ &  0.010 & $ \pm 3.88 \% $ \nl
  5 &  2 &  3 & $   3.21 \times 10^{   -15} $ & -1.055 & $ \pm 3.43 \% $ &  0.002 & $ \pm 0.97 \% $ \nl
  5 &  3 &  3 & $   5.83 \times 10^{   -15} $ & -1.267 & $ \pm 3.47 \% $ & -0.001 & $ \pm 0.15 \% $ \nl
  5 &  4 &  3 & $   8.39 \times 10^{   -15} $ & -1.375 & $ \pm 1.96 \% $ & -0.007 & $ \pm 0.79 \% $ \nl
\tablenotetext{a}{Sum of transitions to a given level from levels $5<n<\infty$}
\tablenotetext{b}{For densities below $n_{e,2}=n_{e}/10^{2}=1$, $\alpha^{\prime}=\alpha^{\prime}_{4} t^{b_{\alpha^{\prime}}}$; for densities in the range $n_{e}=10^{2}-10^{6}~cm^{-3}$,$\alpha^{\prime}=\alpha^{\prime}_{4} t^{b_{\alpha^{\prime}}}n_{e,2}^{c_{\alpha^{\prime}}}$}

\tablenotetext{c}{Error over temperature range $t=0.5-2.0$ holding $n_{e}$ fixed.}
\tablenotetext{d}{Error over density range $\log n_{e}=2-6$ holding $T_{e}$ fixed.}
\end{planotable}



\makeatletter
\def\jnl@aj{AJ}
\ifx\revtex@jnl\jnl@aj\let\tablebreak=\nl\fi
\makeatother


\begin{planotable}{ccccccrllrr}
\tablecaption{Transition data for He I}
\tablehead{
\colhead{$n_{l}$}           & \colhead{$l_{l}$}    &
\colhead{$g_{S,l}$}         & \colhead{$n_{u}$}    &
\colhead{$l_{u}$}           & \colhead{$g_{S,u}$}   &
\colhead{$\lambda_{ul}$}    & \colhead{$A_{ul}$}    & 
\colhead{$\Omega_{4}$}      & \colhead{b$_{\Omega}$}    &
\colhead{e$_{\Omega}$} \\
\colhead{}           & \colhead{}    &
\colhead{}         & \colhead{}    &
\colhead{}           & \colhead{}   &
\colhead{(\AA)}    & \colhead{$s^{-1}$}    & 
\colhead{}      & \colhead{}    &
\colhead{} 
 }
\startdata
  1 &  0 &  1 &  2 &  0 &  3 &      626. & $    1.13 \times 10^{    -4} $ & $    6.86 \times 10^{    -2} $ & 0.020 & $ \pm  4.09 \% $ \nl
  1 &  0 &  1 &  2 &  0 &  1 &      602. & $    5.13 \times 10^{     1} $ & $    3.60 \times 10^{    -2} $ & 0.180 & $ \pm  2.24 \% $ \nl
  1 &  0 &  1 &  2 &  1 &  3 &      591. & $    1.76 \times 10^{     2} $ & $    2.27 \times 10^{    -2} $ & 0.463 & $ \pm  2.60 \% $ \nl
  1 &  0 &  1 &  2 &  1 &  1 &      584. & $    1.80 \times 10^{     9} $ & $    1.54 \times 10^{    -2} $ & 0.631 & $ \pm  1.88 \% $ \nl
  1 &  0 &  1 &  3 &  0 &  3 &      546. &             ---             & $    1.60 \times 10^{    -2} $ & 0.036 & $ \pm  3.38 \% $ \nl
  1 &  0 &  1 &  3 &  0 &  1 &      542. &             ---             & $    8.75 \times 10^{    -3} $ & 0.118 & $ \pm  6.75 \% $ \nl
  1 &  0 &  1 &  3 &  1 &  3 &      540. &             ---             & $    7.39 \times 10^{    -3} $ & 0.324 & $ \pm  6.37 \% $ \nl
  1 &  0 &  1 &  3 &  1 &  1 &      537. & $    5.66 \times 10^{     8} $ & $    3.72 \times 10^{    -3} $ & 0.906 & $ \pm  9.90 \% $ \nl
  1 &  0 &  1 &  3 &  2 &  3 &      538. &             ---             & $    2.30 \times 10^{    -3} $ & 0.094 & $ \pm  1.15 \% $ \nl
  1 &  0 &  1 &  3 &  2 &  1 &      538. &             ---             & $    4.78 \times 10^{    -3} $ & 0.071 & $ \pm  6.66 \% $ \nl
  1 &  0 &  1 &  4 &  0 &  3 &      526. &             ---             & $    6.90 \times 10^{    -3} $ & 0.228 & $ \pm  2.90 \% $ \nl
  1 &  0 &  1 &  4 &  0 &  1 &      524. &             ---             & $    3.52 \times 10^{    -3} $ & 0.435 & $ \pm  7.08 \% $ \nl
  1 &  0 &  1 &  4 &  1 &  3 &      524. &             ---             & $    3.68 \times 10^{    -3} $ & 0.539 & $ \pm  5.64 \% $ \nl
  1 &  0 &  1 &  4 &  1 &  1 &      522. & $    2.43 \times 10^{     8} $ & $    2.23 \times 10^{    -3} $ & 1.236 & $ \pm  6.75 \% $ \nl
  1 &  0 &  1 &  4 &  2 &  3 &      523. &             ---             & $    1.37 \times 10^{    -3} $ & 0.495 & $ \pm  3.38 \% $ \nl
  1 &  0 &  1 &  4 &  2 &  1 &      523. &             ---             & $    2.72 \times 10^{    -3} $ & 0.627 & $ \pm  1.96 \% $ \nl
  1 &  0 &  1 &  4 &  3 &  3 &      523. &             ---             & $    4.90 \times 10^{    -4} $ & 0.089 & $ \pm  1.82 \% $ \nl
  1 &  0 &  1 &  4 &  3 &  1 &      523. &             ---             & $    7.59 \times 10^{    -4} $ & 0.053 & $ \pm  2.15 \% $ \nl
  1 &  0 &  1 &  5 &  1 &  1 &      516. & $    1.26 \times 10^{     8} $ &             ---             & 0.000 & $ \pm  0.00 \% $ \nl
  2 &  0 &  1 &  2 &  1 &  3 &    35649. &             ---             & $    1.70 \times 10^{     0} $ & 0.094 & $ \pm  5.72 \% $ \nl
  2 &  0 &  1 &  2 &  1 &  1 &    20589. & $    1.97 \times 10^{     6} $ & $    1.85 \times 10^{     1} $ & 0.719 & $ \pm 18.44 \% $ \nl
  2 &  0 &  1 &  3 &  0 &  3 &     5905. &             ---             & $    5.67 \times 10^{    -1} $ &-0.375 & $ \pm  2.91 \% $ \nl
  2 &  0 &  1 &  3 &  0 &  1 &     5388. &             ---             & $    6.19 \times 10^{    -1} $ & 0.170 & $ \pm  8.29 \% $ \nl
  2 &  0 &  1 &  3 &  1 &  3 &     5192. &             ---             & $    5.11 \times 10^{    -1} $ &-0.201 & $ \pm  3.66 \% $ \nl
  2 &  0 &  1 &  3 &  1 &  1 &     5017. & $    1.34 \times 10^{     7} $ & $    3.44 \times 10^{    -1} $ & 0.325 & $ \pm  1.64 \% $ \nl
  2 &  0 &  1 &  3 &  2 &  3 &     5051. &             ---             & $    3.33 \times 10^{    -1} $ & 0.285 & $ \pm  1.95 \% $ \nl
  2 &  0 &  1 &  3 &  2 &  1 &     5051. &             ---             & $    1.22 \times 10^{     0} $ & 0.580 & $ \pm  7.16 \% $ \nl
  2 &  0 &  1 &  4 &  0 &  3 &     4169. &             ---             & $    1.96 \times 10^{    -1} $ &-0.190 & $ \pm  0.66 \% $ \nl
  2 &  0 &  1 &  4 &  0 &  1 &     4060. &             ---             & $    1.68 \times 10^{    -1} $ & 0.326 & $ \pm 12.15 \% $ \nl
  2 &  0 &  1 &  4 &  1 &  3 &     4015. &             ---             & $    1.86 \times 10^{    -1} $ &-0.055 & $ \pm  1.02 \% $ \nl
  2 &  0 &  1 &  4 &  1 &  1 &     3966. & $    6.95 \times 10^{     6} $ & $    1.22 \times 10^{    -1} $ & 0.737 & $ \pm  5.46 \% $ \nl
  2 &  0 &  1 &  4 &  2 &  3 &     3979. &             ---             & $    1.24 \times 10^{    -1} $ & 0.513 & $ \pm  1.69 \% $ \nl
  2 &  0 &  1 &  4 &  2 &  1 &     3979. &             ---             & $    3.50 \times 10^{    -1} $ & 0.728 & $ \pm  8.22 \% $ \nl
  2 &  0 &  1 &  4 &  3 &  3 &     3978. &             ---             & $    8.66 \times 10^{    -2} $ & 0.226 & $ \pm  4.93 \% $ \nl
  2 &  0 &  1 &  4 &  3 &  1 &     3978. &             ---             & $    2.58 \times 10^{    -1} $ & 0.663 & $ \pm  2.53 \% $ \nl
  2 &  0 &  1 &  5 &  1 &  1 &     3615. & $    3.80 \times 10^{     6} $ & $    3.23 \times 10^{    -2} $ & 0.602 & $ \pm 18.65 \% $ \nl
  2 &  1 &  1 &  3 &  0 &  3 &     8275. &             ---             & $    1.11 \times 10^{     0} $ &-0.295 & $ \pm  2.91 \% $ \nl
  2 &  1 &  1 &  3 &  0 &  1 &     7283. & $    1.83 \times 10^{     7} $ & $    8.68 \times 10^{    -1} $ & 0.212 & $ \pm  6.95 \% $ \nl
  2 &  1 &  1 &  3 &  1 &  3 &     6940. &             ---             & $    1.18 \times 10^{     0} $ &-0.126 & $ \pm  2.91 \% $ \nl
  2 &  1 &  1 &  3 &  1 &  1 &     6643. &             ---             & $    1.42 \times 10^{     0} $ & 0.638 & $ \pm  2.44 \% $ \nl
  2 &  1 &  1 &  3 &  2 &  3 &     6691. &             ---             & $    8.19 \times 10^{    -1} $ & 0.330 & $ \pm  0.46 \% $ \nl
  2 &  1 &  1 &  3 &  2 &  1 &     6680. & $    6.37 \times 10^{     7} $ & $    3.68 \times 10^{     0} $ & 0.731 & $ \pm  9.75 \% $ \nl
  2 &  1 &  1 &  4 &  0 &  3 &     5226. &             ---             & $    4.19 \times 10^{    -1} $ &-0.109 & $ \pm  1.47 \% $ \nl
  2 &  1 &  1 &  4 &  0 &  1 &     5049. & $    6.77 \times 10^{     6} $ & $    3.15 \times 10^{    -1} $ & 0.320 & $ \pm  9.59 \% $ \nl
  2 &  1 &  1 &  4 &  1 &  3 &     4986. &             ---             & $    4.15 \times 10^{    -1} $ &-0.008 & $ \pm  2.07 \% $ \nl
  2 &  1 &  1 &  4 &  1 &  1 &     4919. &             ---             & $    4.46 \times 10^{    -1} $ & 0.988 & $ \pm  4.65 \% $ \nl
  2 &  1 &  1 &  4 &  2 &  3 &     4931. &             ---             & $    3.14 \times 10^{    -1} $ & 0.492 & $ \pm  1.70 \% $ \nl
  2 &  1 &  1 &  4 &  2 &  1 &     4923. & $    1.99 \times 10^{     7} $ & $    1.16 \times 10^{     0} $ & 0.871 & $ \pm  8.90 \% $ \nl
  2 &  1 &  1 &  4 &  3 &  3 &     4929. &             ---             & $    1.85 \times 10^{    -1} $ & 0.206 & $ \pm  2.27 \% $ \nl
  2 &  1 &  1 &  4 &  3 &  1 &     4929. &             ---             & $    8.13 \times 10^{    -1} $ & 0.702 & $ \pm  4.12 \% $ \nl
  2 &  1 &  1 &  5 &  0 &  1 &     4439. & $    3.27 \times 10^{     6} $ & $    2.99 \times 10^{    -2} $ & 0.584 & $ \pm 14.60 \% $ \nl
  2 &  1 &  1 &  5 &  2 &  1 &     4389. & $    8.99 \times 10^{     6} $ & $    3.87 \times 10^{    -1} $ & 0.586 & $ \pm 14.76 \% $ \nl
  3 &  0 &  1 &  3 &  1 &  3 &   143055. &             ---             & $    3.08 \times 10^{     0} $ &-0.324 & $ \pm  4.89 \% $ \nl
  3 &  0 &  1 &  3 &  1 &  1 &    74405. & $    2.52 \times 10^{     5} $ & $    1.18 \times 10^{     0} $ & 0.927 & $ \pm 11.00 \% $ \nl
  3 &  0 &  1 &  3 &  2 &  3 &    80953. &             ---             & $    1.60 \times 10^{     0} $ & 0.012 & $ \pm  7.86 \% $ \nl
  3 &  0 &  1 &  4 &  0 &  3 &    18432. &             ---             & $    7.36 \times 10^{    -1} $ &-0.588 & $ \pm  2.86 \% $ \nl
  3 &  0 &  1 &  4 &  0 &  1 &    16484. &             ---             & $    2.18 \times 10^{     0} $ & 0.267 & $ \pm 15.39 \% $ \nl
  3 &  0 &  1 &  4 &  1 &  3 &    15765. &             ---             & $    9.01 \times 10^{    -1} $ &-0.424 & $ \pm  2.72 \% $ \nl
  3 &  0 &  1 &  4 &  1 &  1 &    15088. & $    1.41 \times 10^{     6} $ & $    9.79 \times 10^{    -1} $ & 0.594 & $ \pm 10.61 \% $ \nl
  3 &  0 &  1 &  4 &  2 &  3 &    15219. &             ---             & $    7.57 \times 10^{    -1} $ &-0.210 & $ \pm  5.04 \% $ \nl
  3 &  0 &  1 &  4 &  2 &  1 &    15214. &             ---             & $    3.21 \times 10^{     0} $ & 0.441 & $ \pm  8.40 \% $ \nl
  3 &  0 &  1 &  4 &  3 &  3 &    15201. &             ---             & $    1.27 \times 10^{     0} $ & 0.081 & $ \pm  7.87 \% $ \nl
  3 &  0 &  1 &  4 &  3 &  1 &    15201. &             ---             & $    4.16 \times 10^{     0} $ & 0.603 & $ \pm  4.12 \% $ \nl
  3 &  0 &  1 &  5 &  1 &  1 &    11016. & $    9.25 \times 10^{     5} $ & $    2.48 \times 10^{    -1} $ & 1.230 & $ \pm  4.28 \% $ \nl
  3 &  1 &  1 &  4 &  0 &  3 &    24490. &             ---             & $    9.66 \times 10^{    -1} $ &-0.262 & $ \pm  2.95 \% $ \nl
  3 &  1 &  1 &  4 &  0 &  1 &    21137. & $    4.59 \times 10^{     6} $ & $    2.45 \times 10^{     0} $ & 0.635 & $ \pm  9.97 \% $ \nl
  3 &  1 &  1 &  4 &  1 &  3 &    19996. &             ---             & $    1.42 \times 10^{     0} $ &-0.250 & $ \pm  4.68 \% $ \nl
  3 &  1 &  1 &  4 &  1 &  1 &    18951. &             ---             & $    4.21 \times 10^{     0} $ & 0.536 & $ \pm  8.68 \% $ \nl
  3 &  1 &  1 &  4 &  2 &  3 &    19125. &             ---             & $    1.40 \times 10^{     0} $ &-0.008 & $ \pm  6.85 \% $ \nl
  3 &  1 &  1 &  4 &  2 &  1 &    19091. & $    7.12 \times 10^{     6} $ & $    5.75 \times 10^{     0} $ & 0.455 & $ \pm  8.30 \% $ \nl
  3 &  1 &  1 &  4 &  3 &  3 &    19098. &             ---             & $    1.77 \times 10^{     0} $ & 0.186 & $ \pm  5.21 \% $ \nl
  3 &  1 &  1 &  4 &  3 &  1 &    19098. &             ---             & $    8.60 \times 10^{     0} $ & 0.379 & $ \pm 38.16 \% $ \nl
  3 &  1 &  1 &  5 &  0 &  1 &    13414. & $    2.06 \times 10^{     6} $ & $    5.15 \times 10^{    -1} $ & 1.121 & $ \pm  4.82 \% $ \nl
  3 &  1 &  1 &  5 &  2 &  1 &    12972. & $    3.36 \times 10^{     6} $ & $    3.49 \times 10^{     0} $ & 1.146 & $ \pm  4.81 \% $ \nl
  3 &  2 &  1 &  3 &  1 &  1 &   961553. & $    1.53 \times 10^{     2} $ & $    3.30 \times 10^{     1} $ & 1.377 & $ \pm  3.56 \% $ \nl
  3 &  2 &  1 &  4 &  0 &  3 &    23882. &             ---             & $    1.40 \times 10^{     0} $ &-0.309 & $ \pm  1.98 \% $ \nl
  3 &  2 &  1 &  4 &  0 &  1 &    20711. &             ---             & $    2.91 \times 10^{     0} $ & 0.031 & $ \pm  4.15 \% $ \nl
  3 &  2 &  1 &  4 &  1 &  3 &    19589. &             ---             & $    2.52 \times 10^{     0} $ &-0.290 & $ \pm  3.04 \% $ \nl
  3 &  2 &  1 &  4 &  1 &  1 &    18560. & $    2.97 \times 10^{     5} $ & $    4.85 \times 10^{     0} $ & 0.272 & $ \pm  5.18 \% $ \nl
  3 &  2 &  1 &  4 &  2 &  3 &    18753. &             ---             & $    2.54 \times 10^{     0} $ &-0.049 & $ \pm  5.08 \% $ \nl
  3 &  2 &  1 &  4 &  2 &  1 &    18746. &             ---             & $    1.06 \times 10^{     1} $ & 0.237 & $ \pm 22.14 \% $ \nl
  3 &  2 &  1 &  4 &  3 &  3 &    18727. &             ---             & $    3.47 \times 10^{     0} $ & 0.076 & $ \pm  5.53 \% $ \nl
  3 &  2 &  1 &  4 &  3 &  1 &    18701. & $    1.38 \times 10^{     7} $ & $    2.14 \times 10^{     1} $ & 0.696 & $ \pm 10.62 \% $ \nl
  3 &  2 &  1 &  5 &  1 &  1 &    12758. & $    1.27 \times 10^{     5} $ & $    1.18 \times 10^{    -1} $ & 0.977 & $ \pm  2.94 \% $ \nl
  3 &  2 &  1 &  5 &  3 &  1 &    12793. & $    4.54 \times 10^{     6} $ & $    2.85 \times 10^{     1} $ & 0.777 & $ \pm  0.78 \% $ \nl
  4 &  0 &  1 &  4 &  1 &  3 &   361512. &             ---             & $    2.58 \times 10^{     0} $ &-0.488 & $ \pm  4.56 \% $ \nl
  4 &  0 &  1 &  4 &  1 &  1 &   180832. & $    5.83 \times 10^{     4} $ & $    2.17 \times 10^{     1} $ & 0.726 & $ \pm 10.50 \% $ \nl
  4 &  0 &  1 &  4 &  2 &  3 &   198294. &             ---             & $    1.93 \times 10^{     0} $ &-0.275 & $ \pm  8.47 \% $ \nl
  4 &  0 &  1 &  4 &  3 &  3 &   195417. &             ---             & $    1.95 \times 10^{     0} $ &-0.248 & $ \pm  8.60 \% $ \nl
  4 &  0 &  1 &  4 &  3 &  1 &   195417. &             ---             & $    1.52 \times 10^{     1} $ &-0.455 & $ \pm  3.06 \% $ \nl
  4 &  0 &  1 &  5 &  1 &  1 &    33300. & $    2.93 \times 10^{     5} $ & $    6.75 \times 10^{     0} $ & 0.998 & $ \pm  7.61 \% $ \nl
  4 &  1 &  1 &  5 &  0 &  1 &    46062. & $    1.50 \times 10^{     6} $ & $    5.22 \times 10^{     1} $ & 0.826 & $ \pm  6.33 \% $ \nl
  4 &  1 &  1 &  5 &  2 &  1 &    41237. & $    1.52 \times 10^{     6} $ & $    1.57 \times 10^{     2} $ & 0.888 & $ \pm  6.71 \% $ \nl
  4 &  2 &  1 &  4 &  1 &  1 &  2173930. & $    5.66 \times 10^{     1} $ & $    1.55 \times 10^{     2} $ & 1.456 & $ \pm  1.46 \% $ \nl
  4 &  2 &  1 &  4 &  3 &  3 & 18412769. &             ---             & $    7.72 \times 10^{     0} $ &-0.167 & $ \pm  7.89 \% $ \nl
  4 &  2 &  1 &  4 &  3 &  1 & 18389496. &             ---             & $    3.83 \times 10^{     3} $ & 1.308 & $ \pm  4.16 \% $ \nl
  4 &  2 &  1 &  5 &  1 &  1 &    40064. & $    1.63 \times 10^{     5} $ & $    1.05 \times 10^{     1} $ & 0.840 & $ \pm  6.29 \% $ \nl
  4 &  2 &  1 &  5 &  3 &  1 &    40412. & $    2.58 \times 10^{     6} $ & $    3.54 \times 10^{     2} $ & 0.944 & $ \pm  1.21 \% $ \nl
  4 &  3 &  1 &  4 &  3 &  3 &        0. &             ---             & $    1.28 \times 10^{     1} $ &-0.129 & $ \pm  6.69 \% $ \nl
  4 &  3 &  1 &  5 &  2 &  1 &    40559. & $    5.05 \times 10^{     4} $ & $    4.93 \times 10^{     0} $ & 0.944 & $ \pm  1.22 \% $ \nl
  4 &  3 &  1 &  5 &  4 &  1 &    40501. & $    4.25 \times 10^{     6} $ & $    7.48 \times 10^{     2} $ & 0.944 & $ \pm  1.21 \% $ \nl
  5 &  0 &  1 &  5 &  1 &  1 &   358423. & $    1.87 \times 10^{     4} $ & $    1.88 \times 10^{     3} $ & 0.839 & $ \pm  5.65 \% $ \nl
  5 &  2 &  1 &  5 &  1 &  1 &  3999952. & $    2.39 \times 10^{     1} $ & $    1.75 \times 10^{     4} $ & 0.372 & $ \pm  1.68 \% $ \nl
  5 &  2 &  1 &  5 &  3 &  1 & 28407512. &             ---             & $    4.54 \times 10^{     4} $ & 0.477 & $ \pm  3.83 \% $ \nl
  5 &  4 &  1 &  5 &  3 &  1 &        1. &             ---             & $    2.68 \times 10^{     4} $ & 0.403 & $ \pm  2.39 \% $ \nl
  2 &  0 &  3 &  2 &  0 &  1 &    15593. &             ---             & $    2.39 \times 10^{     0} $ &-0.018 & $ \pm  8.35 \% $ \nl
  2 &  0 &  3 &  2 &  1 &  3 &    10833. & $    1.02 \times 10^{     7} $ & $    2.69 \times 10^{     1} $ & 0.812 & $ \pm  2.70 \% $ \nl
  2 &  0 &  3 &  2 &  1 &  1 &     8878. &             ---             & $    9.74 \times 10^{    -1} $ & 0.241 & $ \pm  7.06 \% $ \nl
  2 &  0 &  3 &  3 &  0 &  3 &     4283. &             ---             & $    2.39 \times 10^{     0} $ &-0.045 & $ \pm  4.84 \% $ \nl
  2 &  0 &  3 &  3 &  0 &  1 &     4004. &             ---             & $    3.84 \times 10^{    -1} $ &-0.242 & $ \pm  2.48 \% $ \nl
  2 &  0 &  3 &  3 &  1 &  3 &     3890. & $    9.47 \times 10^{     6} $ & $    1.76 \times 10^{     0} $ & 0.019 & $ \pm  2.70 \% $ \nl
  2 &  0 &  3 &  3 &  1 &  1 &     3800. &             ---             & $    1.57 \times 10^{    -1} $ & 0.176 & $ \pm  5.59 \% $ \nl
  2 &  0 &  3 &  3 &  2 &  3 &     3815. &             ---             & $    2.07 \times 10^{     0} $ & 0.636 & $ \pm  3.70 \% $ \nl
  2 &  0 &  3 &  3 &  2 &  1 &     3815. &             ---             & $    3.08 \times 10^{    -1} $ & 0.034 & $ \pm  1.86 \% $ \nl
  2 &  0 &  3 &  4 &  0 &  3 &     3289. &             ---             & $    7.48 \times 10^{    -1} $ & 0.132 & $ \pm  4.78 \% $ \nl
  2 &  0 &  3 &  4 &  0 &  1 &     3222. &             ---             & $    1.32 \times 10^{    -1} $ &-0.091 & $ \pm  2.18 \% $ \nl
  2 &  0 &  3 &  4 &  1 &  3 &     3189. & $    5.63 \times 10^{     6} $ & $    6.23 \times 10^{    -1} $ & 0.204 & $ \pm  6.27 \% $ \nl
  2 &  0 &  3 &  4 &  1 &  1 &     3165. &             ---             & $    6.39 \times 10^{    -2} $ & 0.450 & $ \pm  5.82 \% $ \nl
  2 &  0 &  3 &  4 &  2 &  3 &     3170. &             ---             & $    7.33 \times 10^{    -1} $ & 0.873 & $ \pm  3.85 \% $ \nl
  2 &  0 &  3 &  4 &  2 &  1 &     3170. &             ---             & $    1.28 \times 10^{    -1} $ & 0.354 & $ \pm  1.48 \% $ \nl
  2 &  0 &  3 &  4 &  3 &  3 &     3169. &             ---             & $    5.05 \times 10^{    -1} $ & 0.501 & $ \pm  2.03 \% $ \nl
  2 &  0 &  3 &  4 &  3 &  1 &     3169. &             ---             & $    5.05 \times 10^{    -2} $ & 0.124 & $ \pm  2.59 \% $ \nl
  2 &  0 &  3 &  5 &  1 &  3 &     2946. & $    3.20 \times 10^{     6} $ & $    5.41 \times 10^{    -2} $ & 0.281 & $ \pm 26.31 \% $ \nl
  2 &  1 &  3 &  2 &  1 &  1 &    48896. &             ---             & $    2.07 \times 10^{     0} $ & 0.411 & $ \pm  6.85 \% $ \nl
  2 &  1 &  3 &  3 &  0 &  3 &     7067. & $    2.78 \times 10^{     7} $ & $    5.70 \times 10^{     0} $ & 0.018 & $ \pm  6.67 \% $ \nl
  2 &  1 &  3 &  3 &  0 &  1 &     6347. &             ---             & $    6.93 \times 10^{    -1} $ &-0.187 & $ \pm  3.10 \% $ \nl
  2 &  1 &  3 &  3 &  1 &  3 &     6077. &             ---             & $    8.06 \times 10^{     0} $ & 0.251 & $ \pm  4.71 \% $ \nl
  2 &  1 &  3 &  3 &  1 &  1 &     5849. &             ---             & $    5.74 \times 10^{    -1} $ & 0.204 & $ \pm  9.03 \% $ \nl
  2 &  1 &  3 &  3 &  2 &  3 &     5877. & $    7.07 \times 10^{     7} $ & $    7.74 \times 10^{     0} $ & 0.867 & $ \pm  5.91 \% $ \nl
  2 &  1 &  3 &  3 &  2 &  1 &     5884. &             ---             & $    8.62 \times 10^{    -1} $ & 0.074 & $ \pm  3.77 \% $ \nl
  2 &  1 &  3 &  4 &  0 &  3 &     4715. & $    9.52 \times 10^{     6} $ & $    1.72 \times 10^{     0} $ & 0.156 & $ \pm  3.80 \% $ \nl
  2 &  1 &  3 &  4 &  0 &  1 &     4582. &             ---             & $    3.17 \times 10^{    -1} $ &-0.089 & $ \pm  1.09 \% $ \nl
  2 &  1 &  3 &  4 &  1 &  3 &     4525. &             ---             & $    2.81 \times 10^{     0} $ & 0.430 & $ \pm  7.75 \% $ \nl
  2 &  1 &  3 &  4 &  1 &  1 &     4469. &             ---             & $    1.94 \times 10^{    -1} $ & 0.491 & $ \pm  6.61 \% $ \nl
  2 &  1 &  3 &  4 &  2 &  3 &     4473. & $    2.46 \times 10^{     7} $ & $    2.72 \times 10^{     0} $ & 0.961 & $ \pm  5.53 \% $ \nl
  2 &  1 &  3 &  4 &  2 &  1 &     4478. &             ---             & $    3.21 \times 10^{    -1} $ & 0.275 & $ \pm  4.95 \% $ \nl
  2 &  1 &  3 &  4 &  3 &  3 &     4477. &             ---             & $    2.18 \times 10^{     0} $ & 0.619 & $ \pm  2.86 \% $ \nl
  2 &  1 &  3 &  4 &  3 &  1 &     4477. &             ---             & $    1.56 \times 10^{    -1} $ & 0.185 & $ \pm  0.91 \% $ \nl
  2 &  1 &  3 &  5 &  0 &  3 &     4122. & $    4.45 \times 10^{     6} $ & $    1.10 \times 10^{    -1} $ & 0.471 & $ \pm 17.08 \% $ \nl
  2 &  1 &  3 &  5 &  2 &  3 &     4027. & $    1.16 \times 10^{     7} $ & $    1.27 \times 10^{     0} $ & 0.472 & $ \pm 17.53 \% $ \nl
  3 &  0 &  3 &  3 &  0 &  1 &    61510. &             ---             & $    2.81 \times 10^{     0} $ &-0.444 & $ \pm  4.84 \% $ \nl
  3 &  0 &  3 &  3 &  1 &  3 &    42955. & $    1.07 \times 10^{     6} $ & $    5.95 \times 10^{     1} $ & 0.000 & $ \pm  0.02 \% $ \nl
  3 &  0 &  3 &  3 &  1 &  1 &    33694. &             ---             & $    1.46 \times 10^{     0} $ &-0.079 & $ \pm  7.17 \% $ \nl
  3 &  0 &  3 &  3 &  2 &  3 &    34952. &             ---             & $    6.38 \times 10^{     1} $ & 0.000 & $ \pm  0.00 \% $ \nl
  3 &  0 &  3 &  3 &  2 &  1 &    34916. &             ---             & $    2.67 \times 10^{     0} $ &-0.235 & $ \pm  3.06 \% $ \nl
  3 &  0 &  3 &  4 &  0 &  3 &    14182. &             ---             & $    7.70 \times 10^{     0} $ & 0.111 & $ \pm 12.08 \% $ \nl
  3 &  0 &  3 &  4 &  0 &  1 &    13000. &             ---             & $    5.37 \times 10^{    -1} $ &-0.509 & $ \pm  1.40 \% $ \nl
  3 &  0 &  3 &  4 &  1 &  3 &    12531. & $    7.09 \times 10^{     5} $ & $    6.67 \times 10^{     0} $ & 0.304 & $ \pm  8.94 \% $ \nl
  3 &  0 &  3 &  4 &  1 &  1 &    12129. &             ---             & $    3.13 \times 10^{    -1} $ &-0.072 & $ \pm  3.71 \% $ \nl
  3 &  0 &  3 &  4 &  2 &  3 &    12200. &             ---             & $    5.74 \times 10^{     0} $ & 0.412 & $ \pm  1.16 \% $ \nl
  3 &  0 &  3 &  4 &  2 &  1 &    12197. &             ---             & $    9.01 \times 10^{    -1} $ &-0.264 & $ \pm  4.25 \% $ \nl
  3 &  0 &  3 &  4 &  3 &  3 &    12189. &             ---             & $    1.25 \times 10^{     1} $ & 0.545 & $ \pm  4.11 \% $ \nl
  3 &  0 &  3 &  4 &  3 &  1 &    12189. &             ---             & $    1.07 \times 10^{     0} $ & 0.100 & $ \pm  8.30 \% $ \nl
  3 &  0 &  3 &  5 &  1 &  3 &     9466. & $    5.69 \times 10^{     5} $ & $    2.72 \times 10^{    -1} $ & 1.206 & $ \pm  2.68 \% $ \nl
  3 &  1 &  3 &  3 &  1 &  1 &   155496. &             ---             & $    3.43 \times 10^{     0} $ & 0.114 & $ \pm  5.83 \% $ \nl
  3 &  1 &  3 &  3 &  2 &  3 &   186220. & $    1.29 \times 10^{     4} $ & $    1.84 \times 10^{     1} $ & 0.717 & $ \pm  9.32 \% $ \nl
  3 &  1 &  3 &  3 &  2 &  1 &   185442. &             ---             & $    6.96 \times 10^{     0} $ &-0.176 & $ \pm  3.70 \% $ \nl
  3 &  1 &  3 &  4 &  0 &  3 &    21128. & $    6.51 \times 10^{     6} $ & $    1.48 \times 10^{     1} $ & 0.463 & $ \pm  5.25 \% $ \nl
  3 &  1 &  3 &  4 &  0 &  1 &    18630. &             ---             & $    9.72 \times 10^{    -1} $ &-0.370 & $ \pm  2.42 \% $ \nl
  3 &  1 &  3 &  4 &  1 &  3 &    17717. &             ---             & $    1.98 \times 10^{     1} $ & 0.251 & $ \pm  5.72 \% $ \nl
  3 &  1 &  3 &  4 &  1 &  1 &    16893. &             ---             & $    1.05 \times 10^{     0} $ &-0.101 & $ \pm  6.94 \% $ \nl
  3 &  1 &  3 &  4 &  2 &  3 &    17007. & $    6.19 \times 10^{     6} $ & $    1.71 \times 10^{     1} $ & 0.330 & $ \pm  5.51 \% $ \nl
  3 &  1 &  3 &  4 &  2 &  1 &    17025. &             ---             & $    2.08 \times 10^{     0} $ &-0.182 & $ \pm  5.55 \% $ \nl
  3 &  1 &  3 &  4 &  3 &  3 &    17009. &             ---             & $    2.98 \times 10^{     1} $ & 0.385 & $ \pm 38.87 \% $ \nl
  3 &  1 &  3 &  4 &  3 &  1 &    17009. &             ---             & $    1.88 \times 10^{     0} $ & 0.177 & $ \pm  4.84 \% $ \nl
  3 &  1 &  3 &  5 &  0 &  3 &    12850. & $    2.73 \times 10^{     6} $ & $    1.89 \times 10^{     0} $ & 1.095 & $ \pm  4.24 \% $ \nl
  3 &  1 &  3 &  5 &  2 &  3 &    11973. & $    3.48 \times 10^{     6} $ & $    8.11 \times 10^{     0} $ & 1.146 & $ \pm  4.25 \% $ \nl
  3 &  2 &  3 &  3 &  1 &  1 &   935906. &             ---             & $    3.61 \times 10^{     0} $ & 0.153 & $ \pm  9.06 \% $ \nl
  3 &  2 &  3 &  3 &  2 &  1 & 33375619. &             ---             & $    6.93 \times 10^{     0} $ & 0.100 & $ \pm  3.86 \% $ \nl
  3 &  2 &  3 &  4 &  0 &  3 &    23865. &             ---             & $    1.45 \times 10^{     1} $ & 0.008 & $ \pm  2.80 \% $ \nl
  3 &  2 &  3 &  4 &  0 &  1 &    20698. &             ---             & $    1.01 \times 10^{     0} $ &-0.289 & $ \pm  1.99 \% $ \nl
  3 &  2 &  3 &  4 &  1 &  3 &    19550. & $    6.46 \times 10^{     5} $ & $    2.28 \times 10^{     1} $ & 0.126 & $ \pm  4.03 \% $ \nl
  3 &  2 &  3 &  4 &  1 &  1 &    18575. &             ---             & $    1.24 \times 10^{     0} $ &-0.009 & $ \pm  7.09 \% $ \nl
  3 &  2 &  3 &  4 &  2 &  3 &    18742. &             ---             & $    3.53 \times 10^{     1} $ & 0.240 & $ \pm 22.42 \% $ \nl
  3 &  2 &  3 &  4 &  2 &  1 &    18735. &             ---             & $    2.72 \times 10^{     0} $ &-0.075 & $ \pm  3.82 \% $ \nl
  3 &  2 &  3 &  4 &  3 &  3 &    18690. & $    1.38 \times 10^{     7} $ & $    6.37 \times 10^{     1} $ & 0.630 & $ \pm 10.88 \% $ \nl
  3 &  2 &  3 &  4 &  3 &  1 &    18716. &             ---             & $    3.24 \times 10^{     0} $ & 0.062 & $ \pm  4.69 \% $ \nl
  3 &  2 &  3 &  5 &  1 &  3 &    12989. & $    2.73 \times 10^{     5} $ & $    8.36 \times 10^{    -1} $ & 0.966 & $ \pm  2.99 \% $ \nl
  3 &  2 &  3 &  5 &  3 &  3 &    12789. & $    4.54 \times 10^{     6} $ & $    8.55 \times 10^{     1} $ & 0.777 & $ \pm  0.78 \% $ \nl
  4 &  0 &  3 &  4 &  0 &  1 &   155979. &             ---             & $    2.27 \times 10^{     0} $ &-0.608 & $ \pm  3.76 \% $ \nl
  4 &  0 &  3 &  4 &  1 &  3 &   108814. & $    2.28 \times 10^{     5} $ & $    6.38 \times 10^{     1} $ & 0.636 & $ \pm 10.87 \% $ \nl
  4 &  0 &  3 &  4 &  1 &  1 &    83798. &             ---             & $    1.08 \times 10^{     0} $ &-0.214 & $ \pm  8.54 \% $ \nl
  4 &  0 &  3 &  4 &  2 &  1 &    87153. &             ---             & $    2.50 \times 10^{     0} $ &-0.352 & $ \pm  7.17 \% $ \nl
  4 &  0 &  3 &  4 &  3 &  3 &    86742. &             ---             & $    4.49 \times 10^{     1} $ &-0.316 & $ \pm  2.87 \% $ \nl
  4 &  0 &  3 &  4 &  3 &  1 &    86742. &             ---             & $    2.06 \times 10^{     0} $ &-0.223 & $ \pm  8.49 \% $ \nl
  4 &  0 &  3 &  5 &  1 &  3 &    28547. & $    1.21 \times 10^{     5} $ & $    4.56 \times 10^{     0} $ & 1.034 & $ \pm  7.57 \% $ \nl
  4 &  1 &  3 &  4 &  1 &  1 &   362822. &             ---             & $    3.58 \times 10^{     0} $ &-0.086 & $ \pm 12.03 \% $ \nl
  4 &  1 &  3 &  4 &  2 &  3 &   438594. & $    4.18 \times 10^{     3} $ & $    8.29 \times 10^{     1} $ & 0.890 & $ \pm  8.62 \% $ \nl
  4 &  1 &  3 &  4 &  2 &  1 &   435388. &             ---             & $    6.83 \times 10^{     0} $ &-0.342 & $ \pm  7.40 \% $ \nl
  4 &  1 &  3 &  4 &  3 &  1 &   425330. &             ---             & $    5.38 \times 10^{     0} $ &-0.281 & $ \pm  7.45 \% $ \nl
  4 &  1 &  3 &  5 &  0 &  3 &    46948. & $    2.02 \times 10^{     6} $ & $    2.48 \times 10^{     2} $ & 0.792 & $ \pm  5.95 \% $ \nl
  4 &  1 &  3 &  5 &  2 &  3 &    37037. & $    1.28 \times 10^{     6} $ & $    2.53 \times 10^{     2} $ & 0.925 & $ \pm  6.96 \% $ \nl
  4 &  2 &  3 &  4 &  1 &  1 &  2086257. &             ---             & $    3.82 \times 10^{     0} $ &-0.145 & $ \pm 10.09 \% $ \nl
  4 &  2 &  3 &  4 &  2 &  1 & 50103835. &             ---             & $    7.66 \times 10^{     0} $ &-0.127 & $ \pm  8.77 \% $ \nl
  4 &  2 &  3 &  4 &  3 &  3 & 13446410. &             ---             & $    7.57 \times 10^{     3} $ & 1.355 & $ \pm  3.74 \% $ \nl
  4 &  2 &  3 &  4 &  3 &  1 & 13464625. &             ---             & $    7.79 \times 10^{     0} $ &-0.173 & $ \pm  7.65 \% $ \nl
  4 &  2 &  3 &  5 &  1 &  3 &    42445. & $    3.27 \times 10^{     5} $ & $    8.31 \times 10^{     1} $ & 0.810 & $ \pm  6.04 \% $ \nl
  4 &  2 &  3 &  5 &  3 &  3 &    40379. & $    2.58 \times 10^{     6} $ & $    1.06 \times 10^{     3} $ & 0.944 & $ \pm  1.21 \% $ \nl
  4 &  3 &  3 &  4 &  1 &  1 &  2468778. &             ---             & $    4.02 \times 10^{     0} $ &-0.185 & $ \pm  9.49 \% $ \nl
  4 &  3 &  3 &  5 &  2 &  3 &    40575. & $    5.05 \times 10^{     4} $ & $    1.48 \times 10^{     1} $ & 0.944 & $ \pm  1.22 \% $ \nl
  4 &  3 &  3 &  5 &  4 &  3 &    40501. & $    4.25 \times 10^{     6} $ & $    2.24 \times 10^{     3} $ & 0.944 & $ \pm  1.21 \% $ \nl
  5 &  0 &  3 &  5 &  1 &  3 &   220265. & $    7.02 \times 10^{     4} $ & $    3.39 \times 10^{     3} $ & 0.925 & $ \pm  5.23 \% $ \nl
  5 &  1 &  3 &  5 &  2 &  3 &   862068. & $    1.51 \times 10^{     3} $ & $    2.37 \times 10^{     4} $ & 0.604 & $ \pm  4.96 \% $ \nl
  5 &  2 &  3 &  5 &  3 &  3 & 22121278. &             ---             & $    1.22 \times 10^{     5} $ & 0.440 & $ \pm  3.29 \% $ \nl
  5 &  4 &  3 &  5 &  3 &  3 &        1. &             ---             & $    8.04 \times 10^{     4} $ & 0.403 & $ \pm  2.39 \% $ \nl

\end{planotable}



\makeatletter
\def\jnl@aj{AJ}
\ifx\revtex@jnl\jnl@aj\let\tablebreak=\nl\fi
\makeatother


\begin{planotable}{cccccrrr}
\tablecaption{He I Emission Line Intensities ($n_{e}=10^{2}~cm^{-3}$)\tablenotemark{a}}
\tablehead{
\colhead{$T=$}           & \colhead{5000}    &
\colhead{10,000}         &
\colhead{20,000}           & \colhead{$a_{j}$}   &
\colhead{$b_{j}$}    & \colhead{$c_{j}$} &
\colhead{$e_{j}$}  }  
\startdata
$ 4 \pi j_{H \beta}/n_{e}n_{H^{+}} $\tablenotemark{b,c} & $   2.20 \times 10^{   -25} $ & $   1.23 \times 10^{   -25} $ & $   6.58 \times 10^{   -26} $ & $   1.37 \times 10^{   -25} $ & -0.982 & -0.104 & $ \pm   0.66\% $ \nl 
$ 4 \pi j_{4686}/n_{e}n_{He^{++}} $\tablenotemark{b,c} & $   3.02 \times 10^{   -24} $ & $   1.52 \times 10^{   -24} $ & $   7.29 \times 10^{   -25} $ & $   1.65 \times 10^{   -24} $ & -1.118 & -0.087 & $ \pm   0.63\% $ \nl 
$ 4 \pi j_{Br \gamma}/n_{e}n_{H^{+}} $\tablenotemark{b,c} & $   2.13 \times 10^{   -25} $ & $   1.15 \times 10^{   -25} $ & $   5.82 \times 10^{   -26} $ & $   1.31 \times 10^{   -25} $ & -1.077 & -0.133 & $ \pm   0.72\% $ \nl 
$ 4 \pi j_{4473}/n_{e}n_{He^{+}} $\tablenotemark{b} & $   1.18 \times 10^{   -25} $ & $   6.16 \times 10^{   -26} $ & $   3.11 \times 10^{   -26} $ & $   6.68 \times 10^{   -26} $ & -1.048 & -0.080 & $ \pm   0.65\% $ \nl 
$ j_{   2946}/j_{4473} $ &  0.356 &  0.438 &  0.540 & $   2.91 \times 10^{   -26} $ & -0.740 & -0.076 &  $ \pm   0.65\% $ \nl 
$ j_{   3189}/j_{4473} $ &  0.745 &  0.913 &  1.136 & $   5.90 \times 10^{   -26} $ & -0.707 & -0.048 &  $ \pm   0.57\% $ \nl 
$ j_{   3615}/j_{4473} $ &  0.094 &  0.110 &  0.127 & $   7.60 \times 10^{   -27} $ & -0.864 & -0.115 &  $ \pm   0.52\% $ \nl 
$ j_{   3890}/j_{4473} $ &  1.864 &  2.231 &  2.742 & $   1.41 \times 10^{   -25} $ & -0.709 & -0.025 &  $ \pm   0.58\% $ \nl 
$ j_{   3966}/j_{4473} $ &  0.193 &  0.226 &  0.261 & $   1.53 \times 10^{   -26} $ & -0.845 & -0.097 &  $ \pm   0.52\% $ \nl 
$ j_{   4027}/j_{4473} $ &  0.450 &  0.464 &  0.463 & $   3.32 \times 10^{   -26} $ & -1.098 & -0.148 &  $ \pm   0.75\% $ \nl 
$ j_{   4122}/j_{4473} $\tablenotemark{*} &  0.028 &  0.040 &  0.059 & $   2.39 \times 10^{   -27} $ & -0.405 &  0.024 &  $ \pm   0.53\% $ \nl 
$ j_{   4389}/j_{4473} $ &  0.120 &  0.123 &  0.121 & $   8.83 \times 10^{   -27} $ & -1.121 & -0.156 &  $ \pm   0.53\% $ \nl 
$ j_{   4439}/j_{4473} $ &  0.013 &  0.017 &  0.023 & $   1.06 \times 10^{   -27} $ & -0.547 & -0.011 &  $ \pm   0.51\% $ \nl 
$ j_{   4473}/j_{4473} $ &  1.000 &  1.000 &  1.000 & $   6.68 \times 10^{   -26} $ & -1.048 & -0.080 &  $ \pm   0.65\% $ \nl 
$ j_{   4715}/j_{4473} $ &  0.075 &  0.105 &  0.164 & $   5.52 \times 10^{   -27} $ & -0.230 &  0.155 &  $ \pm   0.88\% $ \nl 
$ j_{   4923}/j_{4473} $ &  0.269 &  0.266 &  0.257 & $   1.85 \times 10^{   -26} $ & -1.125 & -0.124 &  $ \pm   0.52\% $ \nl 
$ j_{   5017}/j_{4473} $ &  0.496 &  0.567 &  0.646 & $   3.79 \times 10^{   -26} $ & -0.859 & -0.083 &  $ \pm   0.52\% $ \nl 
$ j_{   5049}/j_{4473} $ &  0.031 &  0.041 &  0.057 & $   2.42 \times 10^{   -27} $ & -0.489 &  0.035 &  $ \pm   0.51\% $ \nl 
$ j_{   5877}/j_{4473} $ &  2.989 &  2.743 &  2.591 & $   1.73 \times 10^{   -25} $ & -1.091 & -0.024 &  $ \pm   0.76\% $ \nl 
$ j_{   6680}/j_{4473} $ &  0.857 &  0.778 &  0.702 & $   5.25 \times 10^{   -26} $ & -1.202 & -0.091 &  $ \pm   0.51\% $ \nl 
$ j_{   7067}/j_{4473} $ &  0.358 &  0.481 &  0.713 & $   2.65 \times 10^{   -26} $ & -0.340 &  0.114 &  $ \pm   0.68\% $ \nl 
$ j_{   7283}/j_{4473} $ &  0.114 &  0.146 &  0.197 & $   8.88 \times 10^{   -27} $ & -0.547 &  0.015 &  $ \pm   0.52\% $ \nl 
$ j_{   9466}/j_{4473} $ &  0.020 &  0.024 &  0.030 & $   1.61 \times 10^{   -27} $ & -0.740 & -0.076 &  $ \pm   0.65\% $ \nl 
$ j_{  10833}/j_{4473} $ &  4.298 &  5.515 &  7.895 & $   2.94 \times 10^{   -25} $ & -0.364 &  0.144 &  $ \pm   0.98\% $ \nl 
$ j_{  11973}/j_{4473} $ &  0.045 &  0.047 &  0.047 & $   3.35 \times 10^{   -27} $ & -1.098 & -0.148 &  $ \pm   0.76\% $ \nl 
$ j_{  12531}/j_{4473} $ &  0.024 &  0.029 &  0.036 & $   1.89 \times 10^{   -27} $ & -0.707 & -0.048 &  $ \pm   0.57\% $ \nl 
$ j_{  12789}/j_{4473} $ &  0.174 &  0.152 &  0.129 & $   1.08 \times 10^{   -26} $ & -1.331 & -0.141 &  $ \pm   0.69\% $ \nl 
$ j_{  12793}/j_{4473} $ &  0.058 &  0.051 &  0.043 & $   3.60 \times 10^{   -27} $ & -1.331 & -0.141 &  $ \pm   0.52\% $ \nl 
$ j_{  12972}/j_{4473} $ &  0.015 &  0.016 &  0.015 & $   1.12 \times 10^{   -27} $ & -1.121 & -0.156 &  $ \pm   0.53\% $ \nl 
$ j_{  15088}/j_{4473} $ &  0.010 &  0.012 &  0.014 & $   8.14 \times 10^{   -28} $ & -0.845 & -0.097 &  $ \pm   0.52\% $ \nl 
$ j_{  17007}/j_{4473} $ &  0.066 &  0.066 &  0.066 & $   4.42 \times 10^{   -27} $ & -1.048 & -0.080 &  $ \pm   0.65\% $ \nl 
$ j_{  18690}/j_{4473} $ &  0.440 &  0.360 &  0.295 & $   2.40 \times 10^{   -26} $ & -1.332 & -0.078 &  $ \pm   0.55\% $ \nl 
$ j_{  18701}/j_{4473} $ &  0.146 &  0.120 &  0.097 & $   8.16 \times 10^{   -27} $ & -1.364 & -0.100 &  $ \pm   0.51\% $ \nl 
$ j_{  19091}/j_{4473} $ &  0.025 &  0.025 &  0.024 & $   1.71 \times 10^{   -27} $ & -1.125 & -0.123 &  $ \pm   0.52\% $ \nl 
$ j_{  19550}/j_{4473} $ &  0.014 &  0.017 &  0.021 & $   1.10 \times 10^{   -27} $ & -0.707 & -0.048 &  $ \pm   0.57\% $ \nl 
$ j_{  20589}/j_{4473} $ &  0.629 &  0.670 &  0.721 & $   4.35 \times 10^{   -26} $ & -0.919 & -0.054 &  $ \pm   0.52\% $ \nl 
$ j_{  21128}/j_{4473} $ &  0.011 &  0.016 &  0.025 & $   8.43 \times 10^{   -28} $ & -0.230 &  0.155 &  $ \pm   0.88\% $ \nl 
\tablenotetext{a}{Fits given by ${\rm 4 \pi j_{line}/n_{e}n_{He^{+}}=a_{j}t^{b_{j}}\exp(c_{j}/t)}$, with maximum error $\pm e_{j} \%$ over interval $t=0.5-2.0$.}
\tablenotetext{b}{Emission coefficient in units ${\rm ergs~s^{-1} cm^{3}}$}
\tablenotetext{c}{From Storey \& Hummer (1995)}
\tablenotetext{*}{Flux may be enhanced more than 5 \% for $t \geq 1.5$ due to collisional excitation to $n=5$ levels. See \S 3.1}
\end{planotable}



\makeatletter
\def\jnl@aj{AJ}
\ifx\revtex@jnl\jnl@aj\let\tablebreak=\nl\fi
\makeatother


\begin{planotable}{cccccrrr}
\tablecaption{He I Emission Line Intensities ($n_{e}=10^{4}~cm^{-3}$)\tablenotemark{a}}
\tablehead{
\colhead{$T=$}           & \colhead{5000}    &
\colhead{10,000}         &
\colhead{20,000}           & \colhead{$a_{j}$}   &
\colhead{$b_{j}$}    & \colhead{$c_{j}$} &
\colhead{$e_{j}$}  }
\startdata
$ 4 \pi j_{H \beta}/n_{e}n_{H^{+}} $\tablenotemark{b,c} & $   2.22 \times 10^{   -25} $ & $   1.24 \times 10^{   -25} $ & $   6.59 \times 10^{   -26} $ & $   1.36 \times 10^{   -25} $ & -0.979 & -0.095 & $ \pm   0.69\% $ \nl 
$ 4 \pi j_{4686}/n_{e}n_{He^{++}} $\tablenotemark{b,c} & $   2.96 \times 10^{   -24} $ & $   1.49 \times 10^{   -24} $ & $   7.22 \times 10^{   -25} $ & $   1.63 \times 10^{   -24} $ & -1.110 & -0.086 & $ \pm   0.64\% $ \nl 
$ 4 \pi j_{Br \gamma}/n_{e}n_{H^{+}} $\tablenotemark{b,c} & $   2.14 \times 10^{   -25} $ & $   1.15 \times 10^{   -25} $ & $   5.81 \times 10^{   -26} $ & $   1.31 \times 10^{   -25} $ & -1.078 & -0.127 & $ \pm   0.70\% $ \nl 
$ 4 \pi j_{4473}/n_{e}n_{He^{+}} $\tablenotemark{b} & $   1.18 \times 10^{   -25} $ & $   6.52 \times 10^{   -26} $ & $   5.07 \times 10^{   -26} $ & $   3.36 \times 10^{   -26} $ &  0.092 &  0.665 & $ \pm   3.72\% $ \nl 
$ j_{   2946}/j_{4473} $\tablenotemark{\dag} &  0.357 &  0.419 &  0.359 & $   2.62 \times 10^{   -26} $ & -0.563 &  0.043 &  $ \pm   0.88\% $ \nl 
$ j_{   3189}/j_{4473} $ &  0.748 &  0.931 &  0.997 & $   4.13 \times 10^{   -26} $ &  0.011 &  0.385 &  $ \pm   1.60\% $ \nl 
$ j_{   3615}/j_{4473} $\tablenotemark{*} &  0.094 &  0.104 &  0.078 & $   7.61 \times 10^{   -27} $ & -0.864 & -0.111 &  $ \pm   0.52\% $ \nl 
$ j_{   3890}/j_{4473} $ &  1.891 &  2.555 &  2.953 & $   1.14 \times 10^{   -25} $ &  0.129 &  0.380 &  $ \pm   1.85\% $ \nl 
$ j_{   3966}/j_{4473} $ &  0.194 &  0.219 &  0.187 & $   1.28 \times 10^{   -26} $ & -0.516 &  0.113 &  $ \pm   0.55\% $ \nl 
$ j_{   4027}/j_{4473} $\tablenotemark{\dag} &  0.450 &  0.440 &  0.284 & $   3.31 \times 10^{   -26} $ & -1.097 & -0.144 &  $ \pm   0.76\% $ \nl 
$ j_{   4122}/j_{4473} $\tablenotemark{\dag} &  0.028 &  0.037 &  0.036 & $   2.36 \times 10^{   -27} $ & -0.391 &  0.037 &  $ \pm   0.52\% $ \nl 
$ j_{   4389}/j_{4473} $\tablenotemark{*} &  0.120 &  0.116 &  0.074 & $   8.81 \times 10^{   -27} $ & -1.120 & -0.151 &  $ \pm   0.53\% $ \nl 
$ j_{   4439}/j_{4473} $ &  0.013 &  0.016 &  0.014 & $   1.05 \times 10^{   -27} $ & -0.546 & -0.007 &  $ \pm   0.51\% $ \nl 
$ j_{   4473}/j_{4473} $ &  1.000 &  1.000 &  1.000 & $   3.36 \times 10^{   -26} $ &  0.092 &  0.665 &  $ \pm   3.72\% $ \nl 
$ j_{   4715}/j_{4473} $ &  0.076 &  0.146 &  0.291 & $   4.25 \times 10^{   -27} $ &  1.251 &  0.805 &  $ \pm   5.07\% $ \nl 
$ j_{   4923}/j_{4473} $ &  0.269 &  0.259 &  0.196 & $   1.40 \times 10^{   -26} $ & -0.634 &  0.192 &  $ \pm   0.60\% $ \nl 
$ j_{   5017}/j_{4473} $ &  0.498 &  0.572 &  0.503 & $   3.19 \times 10^{   -26} $ & -0.432 &  0.157 &  $ \pm   0.55\% $ \nl 
$ j_{   5049}/j_{4473} $ &  0.031 &  0.045 &  0.059 & $   1.84 \times 10^{   -27} $ &  0.381 &  0.479 &  $ \pm   0.72\% $ \nl 
$ j_{   5877}/j_{4473} $ &  2.966 &  2.933 &  3.118 & $   8.42 \times 10^{   -26} $ &  0.304 &  0.821 &  $ \pm   3.24\% $ \nl 
$ j_{   6680}/j_{4473} $ &  0.849 &  0.775 &  0.554 & $   4.17 \times 10^{   -26} $ & -0.712 &  0.192 &  $ \pm   0.56\% $ \nl 
$ j_{   7067}/j_{4473} $ &  0.398 &  0.943 &  1.525 & $   6.49 \times 10^{   -26} $ &  0.323 & -0.053 &  $ \pm   4.15\% $ \nl 
$ j_{   7283}/j_{4473} $ &  0.118 &  0.196 &  0.236 & $   1.19 \times 10^{   -26} $ & -0.012 &  0.072 &  $ \pm   0.81\% $ \nl 
$ j_{   9466}/j_{4473} $\tablenotemark{*} &  0.020 &  0.023 &  0.020 & $   1.45 \times 10^{   -27} $ & -0.563 &  0.043 &  $ \pm   0.88\% $ \nl 
$ j_{  10833}/j_{4473} $ & 12.910 & 31.393 & 40.743 & $   3.58 \times 10^{   -24} $ & -0.383 & -0.558 &  $ \pm   2.83\% $ \nl 
$ j_{  11973}/j_{4473} $\tablenotemark{\dag} &  0.045 &  0.044 &  0.029 & $   3.34 \times 10^{   -27} $ & -1.097 & -0.144 &  $ \pm   0.76\% $ \nl 
$ j_{  12531}/j_{4473} $ &  0.024 &  0.030 &  0.032 & $   1.32 \times 10^{   -27} $ &  0.011 &  0.385 &  $ \pm   1.60\% $ \nl 
$ j_{  12789}/j_{4473} $\tablenotemark{\dag} &  0.173 &  0.143 &  0.079 & $   1.07 \times 10^{   -26} $ & -1.328 & -0.138 &  $ \pm   0.69\% $ \nl 
$ j_{  12793}/j_{4473} $ &  0.058 &  0.048 &  0.026 & $   3.58 \times 10^{   -27} $ & -1.328 & -0.138 &  $ \pm   0.52\% $ \nl 
$ j_{  12972}/j_{4473} $\tablenotemark{*} &  0.015 &  0.015 &  0.009 & $   1.12 \times 10^{   -27} $ & -1.120 & -0.151 &  $ \pm   0.53\% $ \nl 
$ j_{  15088}/j_{4473} $ &  0.010 &  0.012 &  0.010 & $   6.79 \times 10^{   -28} $ & -0.516 &  0.113 &  $ \pm   0.55\% $ \nl 
$ j_{  17007}/j_{4473} $ &  0.066 &  0.066 &  0.066 & $   2.22 \times 10^{   -27} $ &  0.092 &  0.665 &  $ \pm   3.72\% $ \nl 
$ j_{  18690}/j_{4473} $ &  0.429 &  0.347 &  0.239 & $   1.59 \times 10^{   -26} $ & -0.663 &  0.353 &  $ \pm   2.33\% $ \nl 
$ j_{  18701}/j_{4473} $ &  0.143 &  0.113 &  0.064 & $   7.32 \times 10^{   -27} $ & -1.192 &  0.006 &  $ \pm   0.53\% $ \nl 
$ j_{  19091}/j_{4473} $ &  0.025 &  0.024 &  0.018 & $   1.29 \times 10^{   -27} $ & -0.634 &  0.192 &  $ \pm   0.60\% $ \nl 
$ j_{  19550}/j_{4473} $ &  0.014 &  0.017 &  0.019 & $   7.72 \times 10^{   -28} $ &  0.011 &  0.385 &  $ \pm   1.60\% $ \nl 
$ j_{  20589}/j_{4473} $ &  0.758 &  1.036 &  0.899 & $   8.35 \times 10^{   -26} $ & -0.708 & -0.212 &  $ \pm   0.65\% $ \nl 
$ j_{  21128}/j_{4473} $ &  0.012 &  0.022 &  0.044 & $   6.49 \times 10^{   -28} $ &  1.251 &  0.805 &  $ \pm   5.07\% $ \nl 
\tablenotetext{a}{Fits given by ${\rm 4 \pi j_{line}/n_{e}n_{He^{+}}=a_{j}t^{b_{j}}\exp(c_{j}/t)}$, with maximum error $\pm e_{j} \%$ over interval $t=0.5-2.0$.}
\tablenotetext{b}{Emission coefficient in units ${\rm ergs~s^{-1} cm^{3}}$}
\tablenotetext{c}{From Storey \& Hummer (1995)}
\tablenotetext{*}{Flux may be enhanced more than 5 \% for $t \geq 1.5$ due to collisional excitation to $n=5$ levels. See \S 3.1}
\tablenotetext{\dag}{Flux may be enhanced more than 25 \% for $t \geq 1.5$ due to collisional excitation to $n=5$ levels. See \S 3.1}
\end{planotable}



\makeatletter
\def\jnl@aj{AJ}
\ifx\revtex@jnl\jnl@aj\let\tablebreak=\nl\fi
\makeatother


\begin{planotable}{cccccrrr}
\tablehead{
\colhead{$T=$}           & \colhead{5000}    &
\colhead{10,000}         &
\colhead{20,000}           & \colhead{$a_{j}$}   &
\colhead{$b_{j}$}    & \colhead{$c_{j}$} &
\colhead{$e_{j}$}  }
\tablecaption{He I Emission Line Intensities ($n_{e}=10^{6}~cm^{-3}$)\tablenotemark{a}}
\startdata
$ 4 \pi j_{H \beta}/n_{e}n_{H^{+}} $\tablenotemark{b,c} & $   2.29 \times 10^{   -25} $ & $   1.26 \times 10^{   -25} $ & $   6.61 \times 10^{   -26} $ & $   1.36 \times 10^{   -25} $ & -0.979 & -0.078 & $ \pm   0.67\% $ \nl 
$ 4 \pi j_{4686}/n_{e}n_{He^{++}} $\tablenotemark{b,c} & $   2.84 \times 10^{   -24} $ & $   1.44 \times 10^{   -24} $ & $   7.04 \times 10^{   -25} $ & $   1.56 \times 10^{   -24} $ & -1.091 & -0.079 & $ \pm   0.64\% $ \nl 
$ 4 \pi j_{Br \gamma}/n_{e}n_{H^{+}} $\tablenotemark{b,c} & $   2.17 \times 10^{   -25} $ & $   1.15 \times 10^{   -25} $ & $   5.76 \times 10^{   -26} $ & $   1.27 \times 10^{   -25} $ & -1.065 & -0.102 & $ \pm   0.70\% $ \nl 
$ 4 \pi j_{4473}/n_{e}n_{He^{+}} $\tablenotemark{b} & $   1.21 \times 10^{   -25} $ & $   6.70 \times 10^{   -26} $ & $   5.58 \times 10^{   -26} $ & $   3.03 \times 10^{   -26} $ &  0.286 &  0.793 & $ \pm   4.02\% $ \nl 
$ j_{   2946 }/j_{4473 } $\tablenotemark{\dag} &  0.361 &  0.417 &  0.335 & $   2.57 \times 10^{   -26 } $ & -0.525 &  0.082 &  $ \pm   0.94\% $ \nl 
$ j_{   3189 }/j_{4473 } $\tablenotemark{*} &  0.759 &  0.943 &  0.981 & $   4.01 \times 10^{   -26 } $ &  0.120 &  0.454 &  $ \pm   1.59\% $ \nl 
$ j_{   3615 }/j_{4473 } $\tablenotemark{*} &  0.096 &  0.104 &  0.072 & $   7.61 \times 10^{   -27 } $ & -0.863 & -0.092 &  $ \pm   0.52\% $ \nl 
$ j_{   3890 }/j_{4473 } $ &  1.929 &  2.677 &  2.991 & $   1.22 \times 10^{   -25 } $ &  0.189 &  0.388 &  $ \pm   2.30\% $ \nl 
$ j_{   3966 }/j_{4473 } $ &  0.197 &  0.219 &  0.178 & $   1.24 \times 10^{   -26 } $ & -0.448 &  0.169 &  $ \pm   0.56\% $ \nl 
$ j_{   4027 }/j_{4473 } $\tablenotemark{\dag} &  0.452 &  0.432 &  0.259 & $   3.29 \times 10^{   -26 } $ & -1.095 & -0.128 &  $ \pm   0.75\% $ \nl 
$ j_{   4122 }/j_{4473 } $\tablenotemark{\dag} &  0.028 &  0.036 &  0.033 & $   2.31 \times 10^{   -27 } $ & -0.382 &  0.055 &  $ \pm   0.52\% $ \nl 
$ j_{   4389 }/j_{4473 } $\tablenotemark{*} &  0.121 &  0.114 &  0.068 & $   8.76 \times 10^{   -27 } $ & -1.117 & -0.134 &  $ \pm   0.53\% $ \nl 
$ j_{   4439 }/j_{4473 } $ &  0.013 &  0.016 &  0.013 & $   1.04 \times 10^{   -27 } $ & -0.542 &  0.009 &  $ \pm   0.51\% $ \nl 
$ j_{   4473 }/j_{4473 } $ &  1.000 &  1.000 &  1.000 & $   3.03 \times 10^{   -26 } $ &  0.286 &  0.793 &  $ \pm   4.02\% $ \nl 
$ j_{   4715 }/j_{4473 } $ &  0.077 &  0.159 &  0.309 & $   5.15 \times 10^{   -27 } $ &  1.263 &  0.726 &  $ \pm   5.81\% $ \nl 
$ j_{   4923 }/j_{4473 } $ &  0.269 &  0.258 &  0.189 & $   1.32 \times 10^{   -26 } $ & -0.524 &  0.269 &  $ \pm   0.62\% $ \nl 
$ j_{   5017 }/j_{4473 } $ &  0.505 &  0.578 &  0.486 & $   3.16 \times 10^{   -26 } $ & -0.363 &  0.201 &  $ \pm   0.77\% $ \nl 
$ j_{   5049 }/j_{4473 } $ &  0.031 &  0.047 &  0.060 & $   1.90 \times 10^{   -27 } $ &  0.471 &  0.503 &  $ \pm   0.78\% $ \nl 
$ j_{   5877 }/j_{4473 } $ &  2.916 &  2.972 &  3.186 & $   8.03 \times 10^{   -26 } $ &  0.484 &  0.908 &  $ \pm   3.12\% $ \nl 
$ j_{   6680 }/j_{4473 } $ &  0.834 &  0.769 &  0.536 & $   4.03 \times 10^{   -26 } $ & -0.607 &  0.247 &  $ \pm   0.56\% $ \nl 
$ j_{   7067 }/j_{4473 } $ &  0.429 &  1.095 &  1.639 & $   9.25 \times 10^{   -26 } $ &  0.182 & -0.232 &  $ \pm   3.91\% $ \nl 
$ j_{   7283 }/j_{4473 } $ &  0.121 &  0.212 &  0.243 & $   1.45 \times 10^{   -26 } $ & -0.051 & -0.018 &  $ \pm   0.85\% $ \nl 
$ j_{   9466 }/j_{4473 } $\tablenotemark{\dag} &  0.020 &  0.023 &  0.019 & $   1.42 \times 10^{   -27 } $ & -0.525 &  0.082 &  $ \pm   0.94\% $ \nl 
$ j_{  10833 }/j_{4473 } $ & 19.346 & 39.911 & 45.367 & $   3.85 \times 10^{   -24 } $ & -0.332 & -0.365 &  $ \pm   3.05\% $ \nl 
$ j_{  11973 }/j_{4473 } $\tablenotemark{\dag} &  0.046 &  0.044 &  0.026 & $   3.32 \times 10^{   -27 } $ & -1.095 & -0.128 &  $ \pm   0.75\% $ \nl 
$ j_{  12531 }/j_{4473 } $\tablenotemark{*} &  0.024 &  0.030 &  0.031 & $   1.28 \times 10^{   -27 } $ &  0.120 &  0.454 &  $ \pm   1.59\% $ \nl 
$ j_{  12789 }/j_{4473 } $\tablenotemark{\dag} &  0.170 &  0.139 &  0.071 & $   1.06 \times 10^{   -26 } $ & -1.318 & -0.125 &  $ \pm   0.69\% $ \nl 
$ j_{  12793 }/j_{4473 } $ &  0.057 &  0.046 &  0.024 & $   3.52 \times 10^{   -27 } $ & -1.318 & -0.125 &  $ \pm   0.52\% $ \nl 
$ j_{  12972 }/j_{4473 } $ &  0.015 &  0.014 &  0.009 & $   1.11 \times 10^{   -27 } $ & -1.117 & -0.134 &  $ \pm   0.53\% $ \nl 
$ j_{  15088 }/j_{4473 } $ &  0.010 &  0.012 &  0.009 & $   6.59 \times 10^{   -28 } $ & -0.448 &  0.169 &  $ \pm   0.56\% $ \nl 
$ j_{  17007 }/j_{4473 } $ &  0.066 &  0.066 &  0.066 & $   2.01 \times 10^{   -27 } $ &  0.286 &  0.794 &  $ \pm   4.02\% $ \nl 
$ j_{  18690 }/j_{4473 } $\tablenotemark{*} &  0.411 &  0.335 &  0.229 & $   1.44 \times 10^{   -26 } $ & -0.510 &  0.443 &  $ \pm   2.57\% $ \nl 
$ j_{  18701 }/j_{4473 } $ &  0.137 &  0.108 &  0.058 & $   6.95 \times 10^{   -27 } $ & -1.129 &  0.042 &  $ \pm   0.54\% $ \nl 
$ j_{  19091 }/j_{4473 } $ &  0.025 &  0.024 &  0.017 & $   1.22 \times 10^{   -27 } $ & -0.524 &  0.269 &  $ \pm   0.62\% $ \nl 
$ j_{  19550 }/j_{4473 } $\tablenotemark{*} &  0.014 &  0.018 &  0.018 & $   7.50 \times 10^{   -28 } $ &  0.120 &  0.454 &  $ \pm   1.59\% $ \nl 
$ j_{  20589 }/j_{4473 } $ &  0.875 &  1.217 &  0.978 & $   1.07 \times 10^{   -25 } $ & -0.755 & -0.269 &  $ \pm   0.72\% $ \nl 
$ j_{  21128 }/j_{4473 } $ &  0.012 &  0.024 &  0.047 & $   7.86 \times 10^{   -28 } $ &  1.263 &  0.726 &  $ \pm   5.81\% $ \nl 
\tablenotetext{a}{Fits given by ${\rm 4 \pi j_{line}/n_{e}n_{He^{+}}=a_{j}t^{b_{j}}\exp(c_{j}/t)}$, with maximum error $\pm e_{j} \%$ over interval $t=0.5-2.0$.}
\tablenotetext{b}{Emission coefficient in units ${\rm ergs~s^{-1} cm^{3}}$}
\tablenotetext{c}{From Storey \& Hummer (1995)}
\tablenotetext{*}{Flux may be enhanced more than 5 \% for $t \geq 1.5$ due to collisional excitation to $n=5$ levels. See \S 3.1}
\tablenotetext{\dag}{Flux may be enhanced more than 25 \% for $t \geq 1.5$ due to collisional excitation to $n=5$ levels. See \S 3.1}
\end{planotable}



\makeatletter
\def\jnl@aj{AJ}
\ifx\revtex@jnl\jnl@aj\let\tablebreak=\nl\fi
\makeatother


\begin{planotable}{ccccccc}
\tablecaption{Populations of metastable $2 ^{3}S$ and $2 ^{1}S$ and resultant collisional 
contributions ($T=20,000~K$)\tablenotemark{a}}
\tablehead{
\colhead{$\log n_{e}=$}           & \multicolumn{2}{c}{2}    &
\multicolumn{2}{c}{4} & \multicolumn{2}{c}{6}  \\
\colhead{} & \colhead{${\rm BSS}^{b}$} & \colhead{${\rm KF}^{c}$} &
\colhead{BSS} & \colhead{KF} &
\colhead{BSS} & \colhead{KF} }
\startdata
$ n(2^{3}S)/n_{He^{+}} $ & $   1.03 \times 10^{    -7} $ & $   1.06 \times 10^{    -7} $ & $   2.04 \times 10^{    -6} $ & $   2.02 \times 10^{    -6} $ & $   2.53 \times 10^{    -6} $ & $   2.46 \times 10^{    -6} $  \nl
$ C/R(   3890.) $ &   0.042 &   0.043 &   0.828 &   0.815 &   1.022 &   0.994  \nl 
$ C/R(   3189.) $ &   0.023 &    ---  &   0.463 &    ---  &   0.571 &    ---   \nl 
$ C/R(   2946.) $ &   0.004 &    ---  &   0.085 &    ---  &   0.105 &    ---   \nl 
$ C/R(   7067.) $ &   0.152 &   0.153 &   3.024 &   2.898 &   3.763 &   3.532  \nl 
$ C/R(   5877.) $ &   0.054 &   0.055 &   1.077 &   1.043 &   1.343 &   1.271  \nl 
$ C/R(   4715.) $ &   0.112 &    ---  &   2.223 &    ---  &   2.777 &    ---   \nl 
$ C/R(   4473.) $ &   0.035 &   0.038 &   0.688 &   0.712 &   0.853 &   0.868  \nl 
$ C/R(   4027.)\tablenotemark{d} $ &    ---  &   0.019 &    ---  &   0.358 &    ---  &   0.436  \nl 
$ C/R(  18690.) $ &   0.018 &    ---  &   0.355 &    ---  &   0.448 &    ---   \nl 
$ n(2^{1}S)/n_{He^{+}} $\tablenotemark{e} & $   7.57 \times 10^{   -14} $ &  --- & $   2.32 \times 10^{   -11} $ &   --- & $   2.71 \times 10^{    -9} $ &   ---  \nl
$ C/R(   5017.) $ &   0.015 &    ---  &   0.287 &    ---  &   0.359 &    ---   \nl 
$ C/R(   3966.) $ &   0.009 &    ---  &   0.175 &    ---  &   0.218 &    ---   \nl 
$ C/R(   7283.) $ &   0.054 &   0.052 &   1.063 &   0.991 &   1.338 &   1.208  \nl 
$ C/R(   6680.) $ &   0.016 &   0.016 &   0.313 &   0.308 &   0.402 &   0.375  \nl 
$ C/R(   4923.) $ &   0.013 &   0.014 &   0.262 &   0.267 &   0.331 &   0.325  \nl 
$ C/R(   4389.) $ &    ---  &   0.011 &    ---  &   0.204 &    ---  &   0.248  \nl 
$ C/R(  18701.) $ &   0.004 &    ---  &   0.082 &    ---  &   0.107 &    ---   \nl 
\tablenotetext{a}{Only lines with collisional corrections,C/R, greater than 1\% over the density range are shown.}
\tablenotetext{b}{Current work.}
\tablenotetext{c}{From Kingdon \& Ferland (1995).}
\tablenotetext{d}{$2 ^{1}S$ population not estimated in Kingdon \& Ferland (1995).}
\tablenotetext{e}{Collisions to $n=5$ not included in current work. See \S3.1.}
\end{planotable}



\makeatletter
\def\jnl@aj{AJ}
\ifx\revtex@jnl\jnl@aj\let\tablebreak=\nl\fi
\makeatother
\begin{planotable}{lcllc}
\tablecaption{Fitting formulae for primordial helium abundance\tablenotemark{a}}
\tablehead{
\colhead{Line}     & \colhead{Formula} & \colhead{$\sigma_{atomic}/f$} &
\colhead{$\sigma_{fit}$} & ${\rm |Max~ Error|}$\tablenotemark{b}}
\startdata
4686\AA & $f=0.0816 t^{0.145}$ & --- & $8.72 \times 10^{-5}$ & $0.25\%$ \nl
6678\AA & $f=2.58 t^{0.249+2.0 \times 10^{-4}n_{e}}$ & 0.013 & $3.05 \times 10^{-3}$ & $0.29 \%$ \nl
4471\AA & $f=2.01 t^{0.127-4.1 \times 10^{-4}n_{e}}$ & 0.013 & $7.67 \times 10^{-3}$ & $0.93 \%$      \nl
5876\AA & $f=0.735 t^{0.230-6.3 \times 10^{-4}n_{e}}$ & 0.015 & $3.06 \times 10^{-3}$ & $1.02 \%$       \nl
\tablenotetext{a}{Helium number abundance is $y=r_{line} f(n_{e},t)$, where
$f=A t^{B_{0}+B_{1}n_{e}}$. Valid over the regime $1.2<t<2.0$ and $1<n_{e}<300 cm^{-3}$.}
\tablenotetext{b}{Maximum difference between calculated value of $f$ and
the fitting function.}
\end{planotable}



\makeatletter
\def\jnl@aj{AJ}
\ifx\revtex@jnl\jnl@aj\let\tablebreak=\nl\fi
\makeatother


\begin{planotable}{ccccccc}
\tablecaption{Quantum defects used for energy levels}
\tablehead{
\colhead{Term=}           & \multicolumn{2}{c}{S}    &
\multicolumn{2}{c}{P} & \multicolumn{2}{c}{D}  \\
\colhead{} & \colhead{$n_{top}$\tablenotemark{a}} & \colhead{$d$\tablenotemark{b}} &
\colhead{$n_{top}$} & \colhead{$d$} &
\colhead{$n_{top}$} & \colhead{$d$} }
\startdata
Singlet & 15  & 0.140  & 20  & -0.012 & 18 & 0.001 \nl
Triplet & 17  & 0.297  & 22  & 0.068  & 21 & 0.003 \nl
\tablenotetext{a}{Top level for which energy levels are given by Martin (1973)}
\tablenotetext{b}{Adopted quantum defect. See Appendix A.}
\end{planotable}


\begin{references}

Almog, Y. \& Netzer, H. 1989, MNRAS, 238, 57 (AN89)

Baldwin, J.A. et al. 1991, ApJ, 374, 580

Berrington, K.B., Burke, P.G., Freitas, L.C.G. \& Kingston, A.E. 1985,
J Phys B, 18, 4135

Berrington, K.B. \& Kingston, A.E. 1987, J Phys B, 20, 6631

Brocklehurst, M. 1971, MNRAS, 153, 471

Brocklehurst, M. 1972, MNRAS, 157, 211

Burgess, A. 1964, ApJ, 139, 776

Burgess, A. 1965, Mem. R. Astonom. Soc., 69, 1

Burgess, A. \& Seaton, M.J. 1960a, MNRAS, 120, 121

Clegg, R.E.S. 1987, MNRAS, 229, 31P

Clegg, R.E.S. \& Harrington, J.P. 1989, MNRAS, 239, 869

Cota, S.A. 1987, PhD Thesis, Ohio State University

Cox, D.P. \& Daltabuit, E. 1971, ApJ, 167, 257

Cunto, W., Mendoza, C., Ochsenbein, F., Zeippen, C. J., 1993, A\&A, 275, L5

Dalgarno, A. \& Kingston, A.E., 1958, Proc. Phys. Soc. London A, 70, 802

Drake, G.W.F. 1979, Phys. Rev A, 19, 1387

Ferland, G.J. 1986, ApJ, 31, L67

Fernley, J. A., Taylor K. T., Seaton M. J. 1987, J. Phys. B Atom. Molec. Phys, 20, 6457

Hata, J. \& Grant, I.P. 1981, J Phys B, 14, 2111

Hyung, S. \& Aller, L.H. 1997, MNRAS, 292, 71

Hyung, S., Aller, L.H., and Feibelman, W.A. 1994a, MNRAS, 269, 975

Hyung, S., Aller, L.H., and Feibelman, W.A. 1994b, ApJS, 93, 465

Izotov, Y.I. \& Thuan, T. X. 1998, ApJ, in press

Janev, R.K., Langer, W.D., Post, D.E., \& Evans, K. 1987. Elementary Processes in Hydrogen-Helium Plasmas (Berlin: Springer)

Kingdon, J. \& Ferland, G.J. 1995, ApJ, 442, 714 (KF95)

Kingdon, J. \& Ferland, G.J. 1996, MNRAS, 282, 723

Kono, A. \& Hattori, S. 1984, Phys Rev A, 29, 2981

Lin, C.D., Johnson, W.R., and Dalgarno, A. 1977, Phys Rev A, 15, 154

Lockyer, J.N. 1868,  Proc. R. Soc., 17, 131

Martin, W.C. 1973, J. Phys. Chem. Ref. Data, 2, 257

Martin, P.G. et al. 1996, BAAS 189, \#106.01

Mathis, J. 1957, ApJ, 125, 318

Mendoza, C. 1983, in IAU Symp. 103, Planetary Nebulae, ed. D. Flower
(Dordrecht: Reidel), 245

Morton, D.C. 1991, ApJS, 77, 119

Olive, K.A., Skillman, E.D., and Steigman, G. 1997a, ApJ, 483,788

Olive, K.A., Skillman, E.D., and Steigman, G. 1997b, ApJ 489, 1006

Osterbrock, D.E. 1989, Astrophysics of Gaseous Nebulae and Active Galactic Nuclei (Mill Valley: University Science Books)

Osterbrock, D.E., Tran, H.D., Veilleux, S., 1992, ApJ, 389, 305

Pagel, B.E.J., Simonson, E.A., Terlevich, R.J., \& Edmunds, M.G. 1992, MNRAS, 255, 325

Peach, G. 1967, Mem. Roy. Astr. Soc., 71, 13

Peimbert, M. \& Torres-Peimbert, S. 1974, ApJ, 193, 327

Pengelly, R.M. \& Seaton, M.J. 1964, MNRAS, 127, 165

Press, W.H., Flanner, B.P., Teukolsky, S.A., and Vetterling, W.T. 1986, Numerical Recipes: The Art of Scientific Computing (New York: Cambridge University Press).

Proga, D., Mikolajewska, J., \& Kenyon, S.J. 1994, MNRAS, 268, 213

Robbins, R.R. 1968, ApJ, 151, 497

Robbins, R.R. \& Bernat, A.P. 1973, Mem. Soc. Roy. Sci. Liege, 6 (5), 263

Robbins, R.R. \& Robinson, E.L. 1971, ApJ, 167, 249

Sasselov, D. \& Goldwirth, D. 1995, ApJL, 444, 5

Sawey, P.M.J. \& Berrington, K.A. 1993, Atomic Data Nucl. Data Tables, 55, 81 (SB93)

Seaton, M.J. 1962, Proc. Phys. Soc., 79, 1105

Skillman, E.D., Terlevich, R.J., Kennicutt, R.C., Garnett, D.R., \&
Terlevich, E. 1994, ApJ,  431, 172

Smits, D.P. 1991a, MNRAS, 248, 193

Smits, D.P. 1991b, MNRAS, 251, 316

Smits, D.P. 1996, MNRAS, 278, 683 (S96)

Storey, P.J. \& Hummer, D.G. 1995, MNRAS, 272, 41

Swartz, D.A. 1994, ApJ, 428, 267

Taylor,I.R., Kingston, A.E., \& Bell, K.L. 1979, J. Phys. B., 12, 3093

Vriens, L. \& Smeet, A.H.M. 1980, Phys. Rev. A, 22, 940.

Walker, T.P., Steigman, G., Schramm, D.N., Olive, K.A., \& Kang,H. 1991, ApJ, 376, 51

Wiese, W.L, Smith, M.W., \& Glennon, B.M. 1966, Atomic Transition Probabilities: NSRDS-NBS 4(1)

\end{references}
\end{document}